\documentclass[
apl,
amsmath,amssymb,
onecolumn,
preprint,
superscriptaddress,
]{revtex4-1}

\usepackage{natbib}
\usepackage{amsmath}
\usepackage{amssymb}
\usepackage{bm}
\usepackage{appendix}
\usepackage{graphicx}
\usepackage{float} 
\usepackage{textcomp}
\usepackage{gensymb}
\usepackage{soul}
\usepackage{xcolor}

\usepackage{dcolumn,color,mathrsfs,titlesec,hyperref,enumerate,subfigure}
\definecolor{linkColor}{rgb}{1,0,0}
\hypersetup{pdfborder={0 0 0},colorlinks=true,urlcolor=linkColor,citecolor=linkColor}

\preprint{}

\begin{document}

\title{The 4D Camera: an 87 kHz direct electron detector for scanning/transmission electron microscopy}


\author{Peter Ercius}
\email{percius@lbl.gov}
\affiliation{National Center for Electron Microscopy, Molecular Foundry, Lawrence Berkeley National Laboratory, 1 Cyclotron Road, Berkeley, CA, USA, 94720}

\author{Ian J. Johnson}
\affiliation{Engineering Division, Lawrence Berkeley National Laboratory, 1 Cyclotron Road, Berkeley, CA, USA, 94720}

\author{Philipp Pelz}
\affiliation{National Center for Electron Microscopy, Molecular Foundry, Lawrence Berkeley National Laboratory, 1 Cyclotron Road, Berkeley, CA, USA, 94720}

\author{Benjamin H. Savitzky}
\affiliation{National Center for Electron Microscopy, Molecular Foundry, Lawrence Berkeley National Laboratory, 1 Cyclotron Road, Berkeley, CA, USA, 94720}

\author{Lauren Hughes}
\affiliation{National Center for Electron Microscopy, Molecular Foundry, Lawrence Berkeley National Laboratory, 1 Cyclotron Road, Berkeley, CA, USA, 94720}

\author{Hamish G. Brown}
\affiliation{National Center for Electron Microscopy, Molecular Foundry, Lawrence Berkeley National Laboratory, 1 Cyclotron Road, Berkeley, CA, USA, 94720}

\author{Steven E. Zeltmann}
\affiliation{Department of Materials Science and Engineering, University of California Berkeley, CA, USA, 94720}

\author{Shang-Lin Hsu}
\affiliation{Department of Materials Science and Engineering, University of California Berkeley, CA, USA, 94720}

\author{Cassio C.S. Pedroso}
\affiliation{Molecular Foundry, Lawrence Berkeley National Laboratory, 1 Cyclotron Road, Berkeley, CA, USA, 94720}

\author{Bruce E. Cohen}
\affiliation{Molecular Foundry, Lawrence Berkeley National Laboratory, 1 Cyclotron Road, Berkeley, CA, USA, 94720}
\affiliation{Division of Molecular Biophysics \& Integrated Bioimaging, Lawrence Berkeley National Laboratory, Berkeley, CA, USA, 94720}

\author{Ramamoorthy Ramesh}
\affiliation{Department of Materials Science and Engineering, University of California Berkeley, CA, USA, 94720}
\affiliation{Materials Science Division, Lawrence Berkeley National Lab, Berkeley, CA, USA, 94720}
\affiliation{Physics Department, University of California, Berkeley, CA, USA, 94720}

\author{David Paul}
\affiliation{National Energy Research Scientific Computing Center, Lawrence Berkeley National Laboratory, Berkeley, CA, USA, 94720}

\author{John M. Joseph}
\affiliation{Engineering Division, Lawrence Berkeley National Laboratory, 1 Cyclotron Road, Berkeley, CA, USA, 94720}

\author{Thorsten Stezelberger}
\affiliation{Engineering Division, Lawrence Berkeley National Laboratory, 1 Cyclotron Road, Berkeley, CA, USA, 94720}

\author{Cory Czarnik}
\affiliation{Gatan, Inc., 5794 W. Las Positas Blvd., Pleasanton, CA, USA, 94588}

\author{Matthew Lent}
\affiliation{Gatan, Inc., 5794 W. Las Positas Blvd., Pleasanton, CA, USA, 94588}

\author{Erin Fong}
\affiliation{Engineering Division, Lawrence Berkeley National Laboratory, 1 Cyclotron Road, Berkeley, CA, USA, 94720}

\author{Jim Ciston}
\affiliation{National Center for Electron Microscopy, Molecular Foundry, Lawrence Berkeley National Laboratory, 1 Cyclotron Road, Berkeley, CA, USA, 94720}

\author{Mary C. Scott}
\affiliation{National Center for Electron Microscopy, Molecular Foundry, Lawrence Berkeley National Laboratory, 1 Cyclotron Road, Berkeley, CA, USA, 94720}
\affiliation{Department of Materials Science and Engineering, University of California Berkeley, CA, USA, 94720}

\author{Colin Ophus}
\affiliation{National Center for Electron Microscopy, Molecular Foundry, Lawrence Berkeley National Laboratory, 1 Cyclotron Road, Berkeley, CA, USA, 94720}

\author{Andrew M. Minor}
\affiliation{National Center for Electron Microscopy, Molecular Foundry, Lawrence Berkeley National Laboratory, 1 Cyclotron Road, Berkeley, CA, USA, 94720}
\affiliation{Department of Materials Science and Engineering, University of California Berkeley, CA, USA, 94720}

\author{Peter Denes}
\affiliation{National Center for Electron Microscopy, Molecular Foundry, Lawrence Berkeley National Laboratory, 1 Cyclotron Road, Berkeley, CA, USA, 94720}

\date{\today}
\begin{abstract}
    We describe the development, operation, and application of the 4D Camera -- a 576 by 576 pixel active pixel sensor for scanning/transmission electron microscopy which operates at 87,000 Hz. The detector generates data at approximately 480 Gbit/s which is captured by dedicated receiver computers with a parallelized software infrastructure that has been implemented to process the resulting 10 - 700 Gigabyte-sized raw datasets. The back illuminated detector provides the ability to detect single electron events at accelerating voltages from 30 - 300 keV. Through electron counting, the resulting sparse data sets are reduced in size by $10 - 300\times$ compared to the raw data, and open-source sparsity-based processing algorithms offer rapid data analysis. The high frame rate allows for large and complex 4D-STEM experiments to be accomplished with typical STEM scanning parameters.
\end{abstract}

\maketitle

\section{Introduction}
Progress in direct electron detectors (DEDs) for transmission electron microscopy (TEM) has recently lead to significant advances in almost every aspect of the field. DEDs provide significantly faster readout, zero dead time (rolling shutter readout), and large improvements in sensitivity compared to traditional charge-coupled device (CCD) cameras. DEDs avoid the conversion process of electrons-to-photons-to-electrons involved in the use of CCDs with a scintillator. This leads to improvements in sensitivity, point spread function (PSF), and detector quantum efficiency (DQE). DEDs have achieved great success in solving 3D biological structures (such as proteins) in the field of Cryo-EM due to their improved sensitivity (to reduce dose) \cite{mcmullan_detective_2009} and speed (to correct motion blur) \cite{grob_ranking_2013, brilot_beam-induced_2012}. Further, \emph{in situ} TEM has greatly benefited from the improved time resolution ($\sim 1$ msec) and use of `rolling readout' mode to eliminate the duty cycle (dead time) of older detector technologies allowing every electron to contribute to the measurement. \cite{park_3d_2015, kim_critical_2020} DEDs are also incorporated into spectrometers for electron energy loss spectroscopy (EELS) and secondary electron microscopes for electron backscatter diffraction (EBSD) with great success. \cite{hart_direct_2017, Susarla2022-ed}

\begin{figure*}
    \centering
   \includegraphics[width=6in]{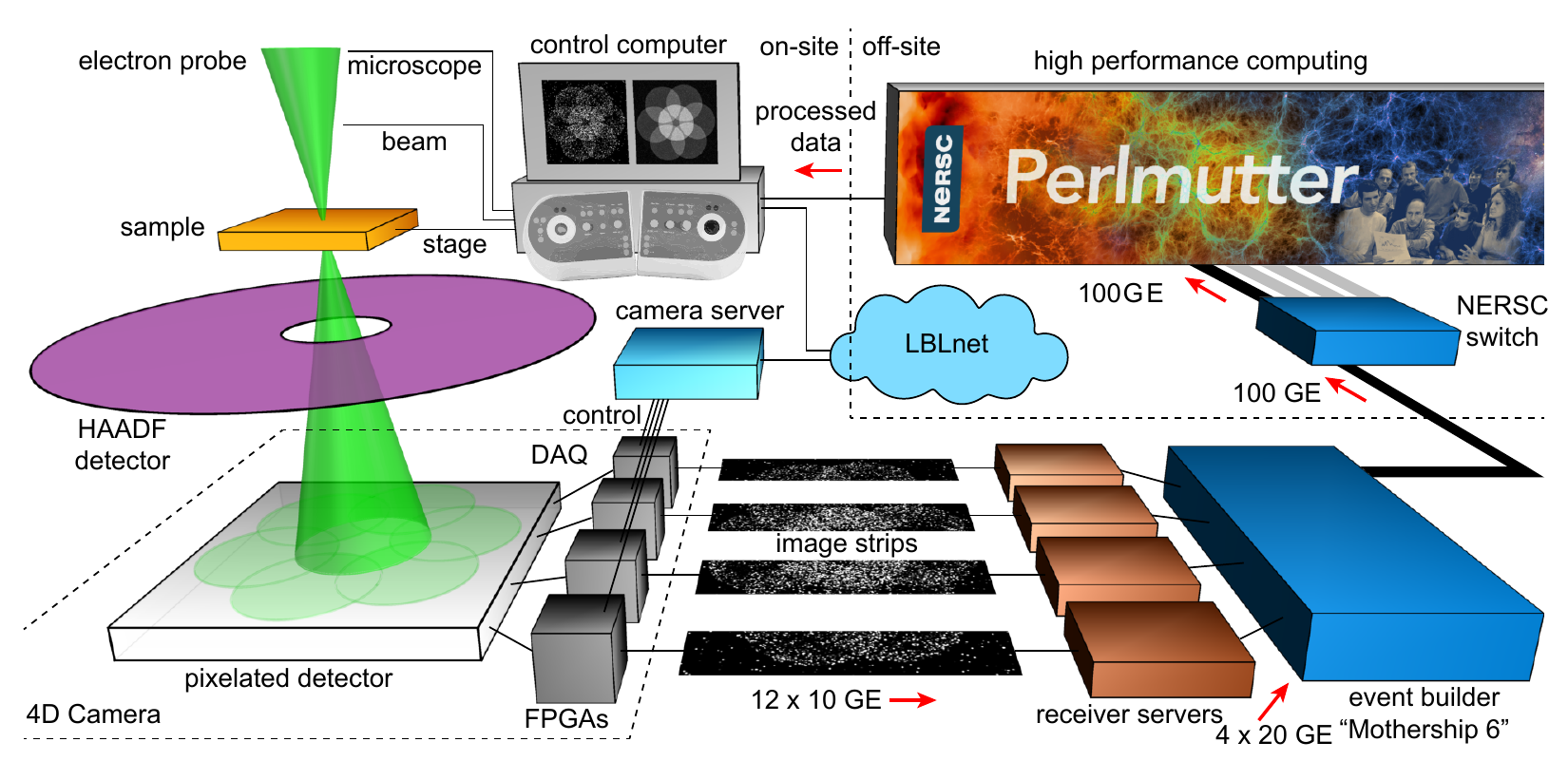}
    \caption[Schematic of the 4D Camera]{A schematic of the 4D Camera detector, data acquisition, and data processing system. Each 1/4 sector of the detector is separately processed by four FPGAs and forwarded to four receiver servers. The receiver servers buffer the data in main memory at the full data rate until a scan is complete. Data is then offloaded to the event builder PC (named ``Mothership 6") which includes a fast connection to the NERSC supercomputer system. Data can be locally or remotely processed for rapid feedback at the microscope.}
    \label{schematic}
\end{figure*}

The application of advanced DEDs for scanning TEM (STEM) allows the user to acquire the two-dimensional convergent beam electron diffraction (CBED) pattern at every scan position in a two-dimensional set of scanning positions. The resulting data set is thus four-dimensional (4D) leading to the common term of 4D-STEM. \cite{ophus_four-dimensional_2019} This is a rapidly evolving field enabling the acquisition of data with traditional STEM image contrast produced by radial, monolithic detectors as well as enabling new imaging modalities \cite{nord2020fast, paterson2020fast}. For example, ptychography has achieved extremely high resolutions \cite{chen_electron_2021} and the ability to image high- and low-scattering elements simultaneously \cite{yang_electron_2017}. Reconstruction of the 3D scattering potential is possible by so-called S-matrix Reconstruction from a focal series of 4D-STEM datasets. \cite{brown_structure_2018, Brown2022-au} Electric field measurements have been shown possible at the atomic level. \cite{hachtel_sub-angstrom_2018, murthy_spatial_2021} Strain and structural measurements at the atomic- to nano-scale are possible due to scanned parallel beam techniques. \cite{hirata_direct_2011, ozdol_strain_2015, zeltmann_patterned_2020} Beam sensitive crystalline materials such as polymers can be investigated with 4D-STEM as well. \cite{panova_diffraction_2019}

The success of DEDs to improve the resolution, sensitivity, flexibility, and efficiency of STEM experiments is truly creating a revolution in atomic- and nano-scale analysis on many fronts. However, most current detectors operate with approximately 1 to 1.5 kHz acquisition rates, which is slower than typical STEM scanning rates of 100 - 1,000 kHz. This can lead to the incorporation of artifacts due to sample drift or charging \cite{jones2015smart, ophus2016correcting}, especially in ptychography where phase information about the sample is found using the overlapping of adjacent probes. kHz acquisition rates also limit the achievable minimum dose applied to a sample, because the operator must reduce the beam current by a factor of 10 to 100 when switching from ADF-STEM imaging, which is used for focusing and aligning, to 4D-STEM in order to maintain the same dose.\cite{oleary_phase_2020} Further, slow acquisition rates limit the use of 4D-STEM for large field-of-view acquisitions and for complex STEM acquisition schemes such as focal series, electron tomography, scan rotation series, \textit{in situ}, and high throughput experiments. Recent successes have achieved excellent results at high frame rates at the cost of pixel binning. \cite{Stroppa2023-vy} The 4D Camera was designed to acquire full frame 4D-STEM data with a probe dwell time of 11 $\mu$s making 4D-STEM highly complementary to traditional STEM detectors.

\section{Camera setup and acquisition system}

The 4D Camera sensor is mounted in a retractable Gatan K3 camera housing, making it immediately deployable on any TEM. It is currently installed on a double aberration-corrected Titan 80-300 named the TEAM 0.5 at the National Center for Electron Microscopy (NCEM) facility within the Molecular Foundry at Lawrence Berkeley National Laboratory (LBNL). Each frame consists of 576 by 576 pixels digitized to 12 bits, which is about 4 Mbit per frame. At 87,000 Hz, the raw data rate is approximately 480 Gbit/s requiring a dedicated acquisition system to capture the full data rate. The full detector system (see Figure \ref{schematic}) includes a set of field programmable gate arrays (FPGAs) for collecting and routing the digitized detector pixel signals and a set of computers for data capture, storage, and analysis. A 1/4 sector of each frame is handled separately by a dedicated FPGA / receiver pair during acquisition. The raw data is accumulated at the full rate into the memory of the four dedicated receiver servers over a high speed network consisting of 48~x~10 Gbit/s connections (12 links per receiver). After the STEM scan is complete, the data is then offloaded to flash memory installed in a fifth event builder server over 4~x~20 Gbit/s network connections.

The fifth server acts as the storage and processing computer for data reduction, post-processing, and interactive analysis and is directly connected to the National Energy Research Computing Center (NERSC) by 100 Gbit fiber providing extra computational and storage capabilities for the large amounts of data generated. The open-source stempy \cite{Avery2022-if} processing software was developed to rapidly reduce the raw data by finding electron strikes, and a 1024 by 1024 probe position data set can be reduced from 700 GB to less than 10 GB in under ten minutes using local or remote processing. Other post-processing such as virtual annular detector (vADF-) STEM image generation is then accomplished using Jupyter notebooks applied directly to the sparse output data. The operator receives rapid feedback at the microscope to determine the quality of each scan even for these extremely large data sets.

\begin{figure*}
    \centering
    \includegraphics[width=6.4in]{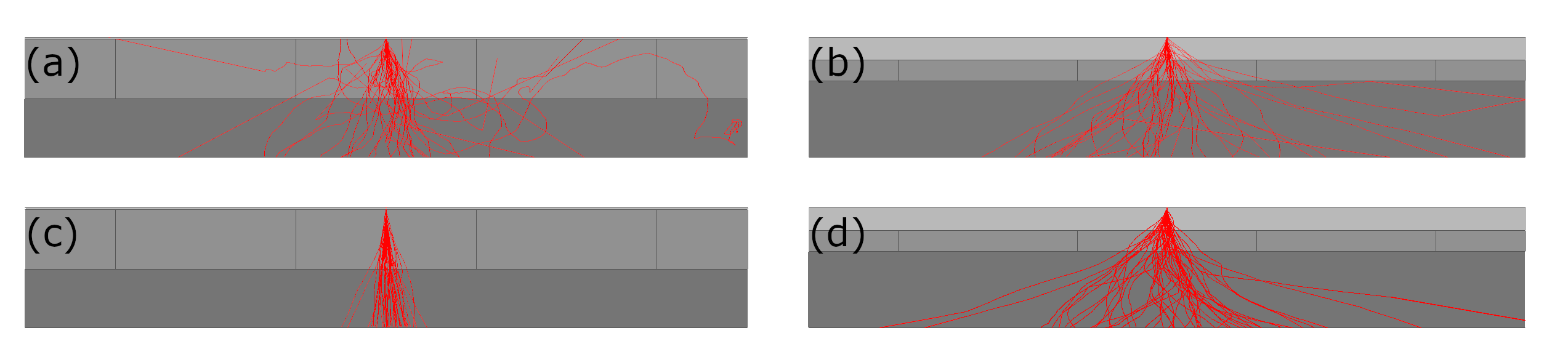}
    \caption[scattering in pixels]{Schematics of (a)-(b) 80 kV and (c)-(d) 300 kV electron scattering (red lines) in 10 $ \mu$m wide APS pixels (vertical gray lines). (a) and (c) show how a thinner inactive layer reduces electron scattering leading to reduced lateral energy deposition compared to a sensor with a thicker inactive layer ((b) and (d)).}
    \label{pixels}
\end{figure*}

\section{Sensor details}

The sensor is a complementary metal oxide semiconductor (CMOS) active pixel sensor (APS) with 10$\mu$m by 10$\mu$m square pixels. APS sensors are designed to capture a few scattering events for each incident electron in the charge collection area while minimizing lateral charge deposition. A schematic of such a CMOS sensor is shown in Figure \ref{pixels} which demonstrates how front-side illuminated sensors with regions of material above the charge collection region broaden the range of energy deposition due to electron scattering. In general, front-side illuminated CMOS detectors are limited to operating at relatively high accelerating voltages (i.e. 300 keV) to minimize this effect and avoid damaging the chip electronics. The 4D Camera sensor was produced by thinning the inactive substrate layer to reduce electron scattering outside the active area. The sensor is back-side illuminated thus exposing the charge collection region (see Figure \ref{pixels} (a) and (c)) first to the beam. The detector was tested at accelerating voltages from $6 - 300$ kV, and full charge deposition is seen at voltages below 30 kV. The sensor used to acquire the data in this manuscript was over-thinned, removing some of the active region making it comparatively more sensitive at lower accelerating voltages.

For typical STEM experiments, the incident electron beam flux is sufficiently low, such that only about 1\% - 2.5\% of the detector pixels will be hit with an electron in each frame; however, it can also be used in ``integration" mode where each pixel intensity is the sum of all charge deposited in a pixel allowing full frame movies to be acquired with high current. Here, we discuss the detector for use in only the sparse counting mode. During the design phase, we considered typical parameters for aberration-corrected STEM operating conditions (60-300 kV, $\sim 10 \mu$s dwell time, 1 - 100 pA beam current) in order to match 4D Camera operation as close as possible to normal operational parameters. Using these parameters, 87,000 Hz was a chosen to match the $\sim$ 10 $\mu$s dwell time used in typical HR-STEM image acquisitions. At such acquisition speeds and beam currents only about 1 electron will strike a pixel and the surrounding nearest- and second-nearest neighbors. The energy deposited in a pixel and the surrounding pixels is then used to determine the initial pixel location of the electron strike.\cite{battaglia_cluster_2009} The nature of electron scattering in a material dictates that the electron on average will deposit energy in a large surrounding area \cite{caswell_high-speed_2009}, leading to blurring called the point spread function (PSF). As shown in Figure \ref{pixels}, thinning the inactive layer and back-side illumination will reduce electron scattering, but the surrounding pixels can still have energy deposited in them. Further, the number of scattering events and amount of energy deposited is a stochastic process leading to a Landau distribution of energy deposition in each pixel.\cite{denes_active_2007} It is thus impractical to use the amount of energy captured in each pixel to differentiate the number of electrons. However, if the flux on the detector is sufficiently low, then we can assume with high confidence that only one electron will strike within any 3 by 3 pixel neighborhood creating a local peak. The pixel with the highest deposited energy is very likely the original strike location of the electron, furthermore sub-pixel locations can be determined using centroiding.\cite{li_influence_2013, ruskin_quantitative_2013} In the simple case, the highest intensity pixel is counted as one electron regardless of energy deposition, and the surrounding pixels are set to zero which removes the PSF blurring. This also leads to suppression of detector dark current noise and the ability to sum frames without adding detector noise.\cite{battaglia_cluster_2009} An added benefit is that the counted data are highly compressible leading to smaller storage and memory requirements and the ability to implement faster computation with algorithms designed to take advantage of this sparse data format.\cite{pelz_real-time_2022} What would be extremely large ($>$ 100 Gbyte) datasets can be compressed $10 - 300\times$ with the ability to fit in the working memory of consumer grade computers and GPUs. The 4D Camera has been in operation since approximately Spring of 2019. The capabilities of the detector for several operating modes is presented and discussed in more detail below.

\section{HR-STEM: virtual detectors and phase contrast imaging}

\begin{figure*}
	\centering
	\includegraphics[width=6.4in]{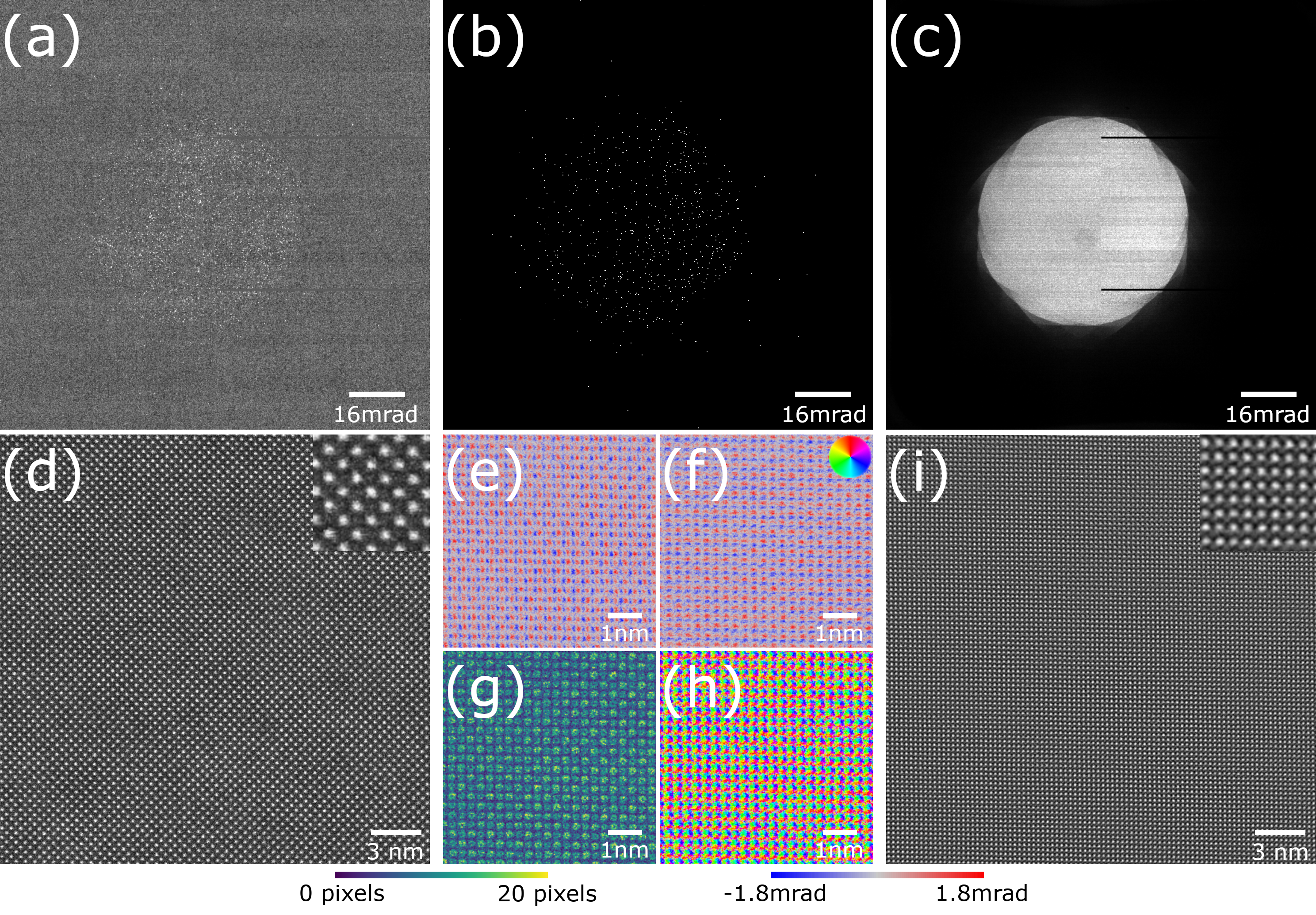}
	\caption[PTO]{HR-STEM imaging of a STO/PTO multi-layer. a) A raw camera frame acquired in $11 \mu s$ from one scan position. b) A single counted frame showing only electron strikes as bright pixels. c) A PACBED pattern generated by summing the counted data. d) The simultaneously acquired ADF-STEM image and magnified inlay. e-h) The CoM$_x$, CoM$_y$, CoM$_r$, CoM$_\theta$ of a small 250 by 250 pixel region of the full scan. i) The DPC image calculated from the CoM$_x$ and CoM$_y$ with magnified inlay. The inlays have a field-of-view of 2.3 nm.}
	\label{PTO}
\end{figure*}

\subsection{STO/PTO heterostructure}

Typical STEM imaging experiments use radially symmetric, monolithic bright field (BF), annular bright field (ABF), and/or annular dark field (ADF) detectors to integrate electrons scattered to different angles in diffraction space. The choice of detectors and their inner/outer angles must be decided before an experiment is conducted. The use of a pixelated detector to capture the full scattering pattern provides the ability to generate traditional STEM image contrast with the flexibility of choice as to the inner/outer angle(s) during post-processing. The 4D Camera provides the ability to acquire a large, continuous range of scattering angles in the same amount of time as a typical STEM experiment ($<$ 15 sec). Once electron-counted, a radial sum algorithm implemented in the sparse domain \cite{Avery2022-if, pelz_real-time_2022} further reduces the 4D data to 3D and allows a user to interactively set detector inner/outer angles for the desired STEM image contrast.

Figure \ref{PTO} shows data acquired at high resolution from a highly converged electron beam of a SrTiO$_3 (20)$ / PbTiO$_3 (20)$ / SrTiO$_3 (20)$ multilayer sample acquired perpendicular to the interfaces (i.e. planview) along the $[100]$ direction. The sample was detached from a substrate and placed on a TEM grid using the process described in \cite{bakaul_single_2016}, and the nominal thickness was 23.64 nm. The STEM was operated at 300 kV with a convergence angle of 30 mrad and a beam current of about 100 pA. The thinned sensor is approximately 30\% efficient at detecting 300 kV electrons, necessitating larger beam currents than should be necessary. Figure \ref{PTO}(a) shows a single raw camera frame. A faint outline of the central circular beam is visible but is obscured by camera dark noise. Figure \ref{PTO}(b) shows the same frame after it has been electron counted where each bright pixel indicates the location of one electron strike. Most electrons are within the central beam although some are scattered outside. Summation of all sparse frames from every scan position produces a position-averaged convergent beam electron diffraction (PACBED) pattern, as shown in Figure \ref{PTO}(c), which can be used to determine sample thickness, composition, and other properties \cite{lebeau2010position, ophus2017nonspectroscopic}. Two dark horizontal rows are due to production defects in this first iteration of the detector, and other horizontal defects are due to non-linear camera noise which are not removed due to the thin detector. Figure \ref{PTO}(d) shows the simultaneously acquired ADF-STEM image where the projected Pb and Ti / O columns show the expected Z-contrast.

\begin{figure*}
	\centering
	\includegraphics[width=3.3in]{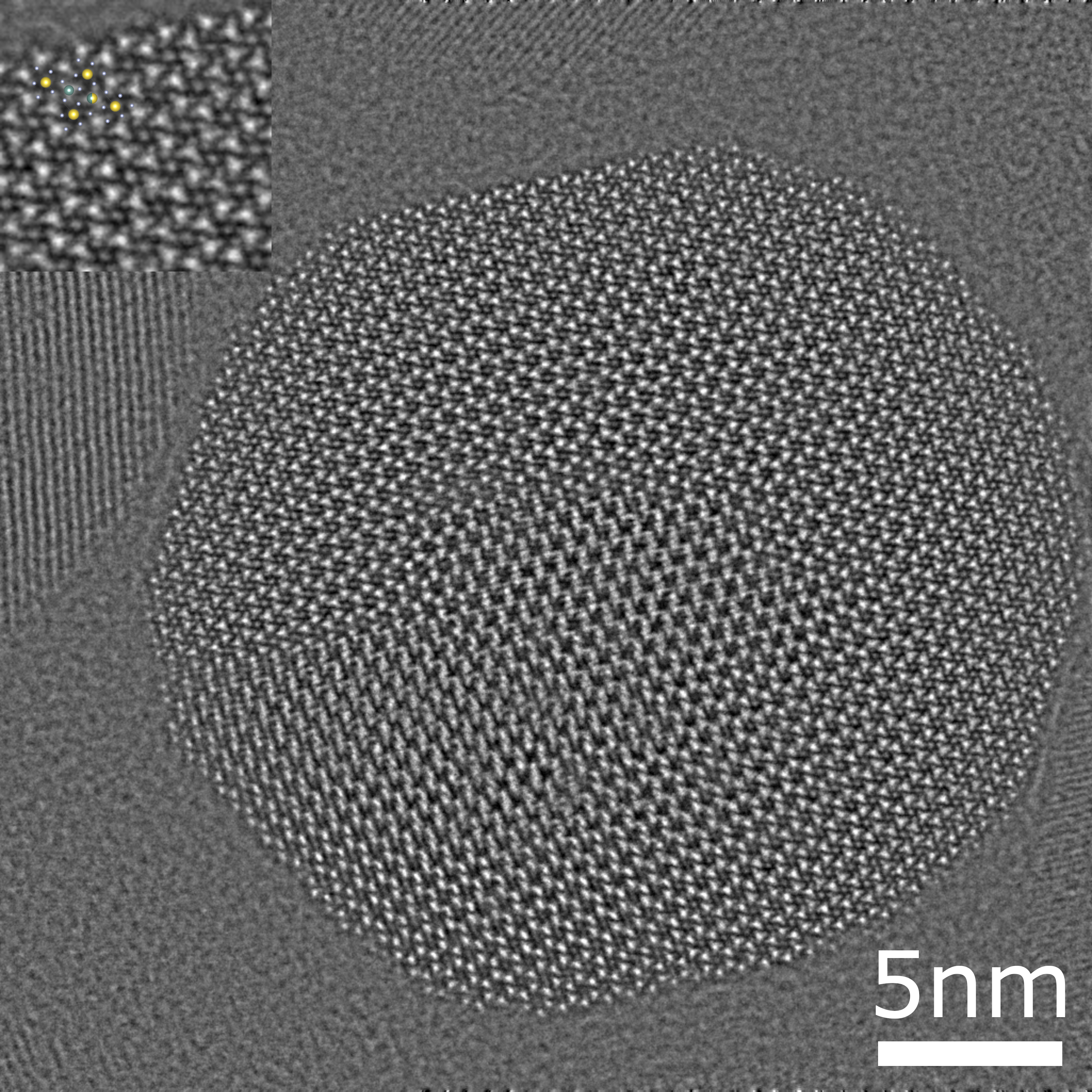}
	\caption[NaYF]{Low-dose HR-STEM DPC phase imaging of a doped NaYF$_4$ nanoparticle. The inset shows a 4 nm by 4 nm magnified region of the edge of the NP where yttrium, sodium, and fluorine columns are visible.}
	\label{NaYF}
\end{figure*}

Phase contrast imaging is a powerful technique which can simultaneously image the weakly and strongly scattering atoms in materials \cite{ophus2016efficient}. It can be implemented using pixelated area detectors \cite{yang_simultaneous_2016} or segmented monolithic detectors \cite{shibata2012differential}, where the measurement of the phase is accomplished using e.g. differential phase contrast (DPC) algorithms.\cite{close_towards_2015, ishizuka_boundary-artifact-free_2017} The center of mass (CoM) of each pattern is calculated giving a measurement of the shift of the central disk or a redistribution of intensity within the disk as it is scanned across the sample.\cite{cao_theory_2018} For sparse 4D Camera data, the CoM calculation is implemented in the sparse domain where only the positions of electrons are considered, leading to a faster calculation compared to normal dense datasets. \cite{pelz_real-time_2022}

The CoM was calculated for each counted frame in Figure \ref{PTO}. Even with the small number of electrons (see Figure \ref{PTO}(b)) the CoM$_x$ and CoM$_y$ signal is quite strong as can be seen in the zoomed in Figure \ref{PTO}(e)-(f), respectively. Atom positions are at the center of each red/blue blob. The radial deviation of the CoM (Figure \ref{PTO}(g)) shows each atomic column with a singularity in the center as the beam passes directly over the column, and the vector angle is plotted in Figure \ref{PTO}(h). Finally, the CoM images are used to calculate the DPC signal resulting in Figure \ref{PTO}(i) over a 25.4 nm field of view with 0.025 nm real space pixel size. In DPC, it is known that the A site and B site columns should have a more similar contrast than presented in ADF-STEM due to the additional phase shift introduced by the oxygen atoms not typically imaged by high angle scattering. In Figure \ref{PTO}i, the pure oxygen columns are visible, but difficult to resolve at this length scale. Other datasets acquired with smaller probe step sizes clearly show the oxygen columns, but we display this medium magnification data set to demonstrate the ability to scan large regions for e.g. polarization measurements. \cite{yadav_observation_2016}


\subsection{Dose Sensitive \texorpdfstring{NaYF$_4$}{NaYF4} nanoparticle}

To show the capabilities of the detector for low-dose imaging of sensitive materials with high- and low-scattering elements, we chose a hexagonal NaYF$_4$ nanoparticle with core and shell doping of 15\% thulium and 20\% gadolinium, respectively. \cite{Pedroso2021-wv} These avalanching upconverting nanoparticles exhibit extremely large amplification factors due to a photo avalanche mechanism and are useful as probes in optical imaging of biological systems and high-density, rewritable 3D photopatterning.\cite{Lee2021-rz, Lee2022-gw} Several questions about the structure of these core/shell nanoparticles remain such as the dispersion of defects and dopants and the core/shell interface which could be answered by detailed high resolution analysis. The material is sensitive to the electron beam \cite{hobbs1973transmission} and contains high- and low-scattering elements with small inter-atomic spacings difficult to resolve.

The NPs are best imaged along the c-axis where mixing of atom species along the projection direction is minimized. Low-dose HAADF-STEM measurements can resolve the yttrium atoms due to their relatively large Z-number, but the other atomic columns are weak scatterers. High resolution 4D-STEM scans were acquired using the 4D Camera with a 200 kV accelerating voltage, 17.1 mrad convergence angle, and each diffraction pattern contained approximately 150 electrons (2 pA beam current). 1024 by 1024 real space probe positions were required to image the entire 26 nm diameter NP in a 30 by 30 nm field of view with 0.029 nm pixel size with a total dose of approximately 1800 $e^-/$ \AA$^2$. This allows the fluorine atomic columns to be resolved, while minimizing sample damage. The DPC signal was extracted from CoM measurements of each diffraction pattern to produce the image shown in Figure \ref{NaYF}. The inset shows the atomic arrangement overlaid on a magnified portion of the top edge of the NP where the hexagonal rings of the yttrium and 50/50 yttrium/sodium columns are seen as the brightest columns. At the center of each bright hexagon is a dimmer sodium column. Triangular arrangements of fluorine columns surround the brighter columns. In the center, some defects due to beam damage are seen, but the edge is fairly stable all around the NP. We found that the NP centers were more easily damaged by the beam which is either due to the thulium dopants in the core or more inelastic scattering due to the increase in projected thickness. The ability to simultaneously image high- and low-scattering elements in a large beam sensitive NP shows the capabilities of the 4D Camera for investigating new classes of materials previously too sensitive to image at high resolution using these techniques.

\subsection{Ptychography}

\begin{figure*}
	\centering
	\includegraphics[width=6.4in]{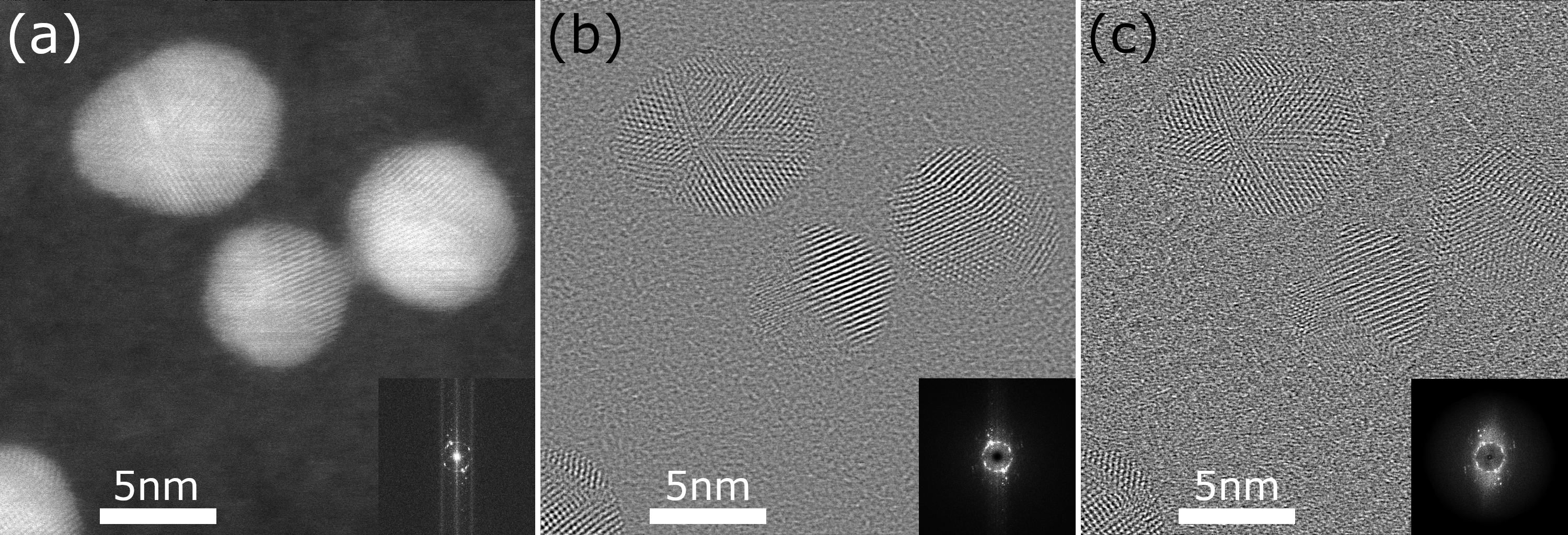}
	\caption[ptychography]{HR-STEM DPC and ptychographic imaging of gold NPs on carbon. (a) ADF-STEM image generated from the sparse 4D data set. (b) Single-pass DPC image created from the sparse 4D data set (c) The single side band ptychographic image from the sparse data set.}
	\label{Au_SSB}
\end{figure*}

DPC is one type of phase contrast imaging that is relatively simple to extract from CoM measurements made by a pixelated detector. It does not however utilize the full range of information possible with such a detector and can be implemented using simple segmented detector with only four channels. \cite{shibata2012differential, bosch_integrated_2016} A more advanced algorithm capable of using the entire scattering distribution is called ptychography. \cite{humphry_ptychographic_2012, jiang_electron_2018} We implemented a ptychographic reconstruction method that takes advantage of the sparse nature of 4D Camera data providing large scale, rapid analysis on a modest GPU. The method is described in more detail in \cite{pelz_real-time_2022} and is based on the single sideband (SSB) algorithm developed previously. \cite{yang_electron_2017, yang_efficient_2015} In brief, the SSB ptychography method takes the Fourier transform of the 4D data hypercube along the probe scanning directions. Summation of the intensity within the double overlap regions of the double reciprocal space data set can then be used to determine the object function which approximates the projected potential of the sample in the weak scattering limit. The speed and sparse data output of the 4D Camera provides the ability to rapidly generate ptychographic reconstructions over a wide field-of-view in less than 10 minutes.

Figure \ref{Au_SSB} demonstrates the ptychographic reconstruction capabilities of the 4D Camera from a 1024 by 1024 4D-STEM scan with 0.03 nm real space pixel sampling of gold nanoparticles on an ultra-thin carbon substrate. The beam accelerating voltage was 80 kV, the convergence angle was 30 mrad, and the beam current was approximately 80 pA. At this accelerating voltage, the detector is approximately 50\% efficient as measured by comparing the measured screen current from the Titan software to the flux of electron strikes incident on the detector. Figure \ref{Au_SSB}(a) shows the virtual ADF-STEM image generated by summing all electron strikes within 46 - 72 mrad showing several approximately 5 nm gold nanoparticles. Figure \ref{Au_SSB}(b) shows the DPC reconstruction using the same technique as used for Figure \ref{PTO}(i) and \ref{NaYF} where the atomic contrast is better resolved and more features are seen in the lightly scattering carbon substrate. Figure \ref{Au_SSB}(c) shows the SSB ptychographic reconstruction from the same data set where more low frequencies and improved atomic contrast is evident with the inset Fourier transforms. Implementation of the SSB algorithm to take advantage of the sparse data output allows a 1024 by 1024 probe position 4D-STEM data set, which is originnally 700 GB in raw form, to be processed by a GPU in less than 1 second. \cite{pelz_real-time_2022}


\section{Parallel beam large field-of-view acquisitions}

Another common experimental mode of 4D-STEM is nanobeam electron diffraction (NBED), where the convergence angle of the STEM probe is reduced to avoid overlap of diffracted beams on the detector.\cite{ophus_four-dimensional_2019} The probe is localized to approximately the nanometer scale and can provide information about crystal phase, orientation, and strain while also providing the ability to generate virtual BF and virtual DF real space images from any desired scattering angles. The speed of the 4D Camera provides the ability to scan over very large fields-of-view reducing the need to carefully choose the region of interest. Full devices or a battery stack can be rapidly investigated with nanometer scale real space pixel sizes over a 10's of micron-sized field of view.

\subsection{LiPON Battery Stack}

\begin{figure*}
    \centering
   \includegraphics[width=2.8in]{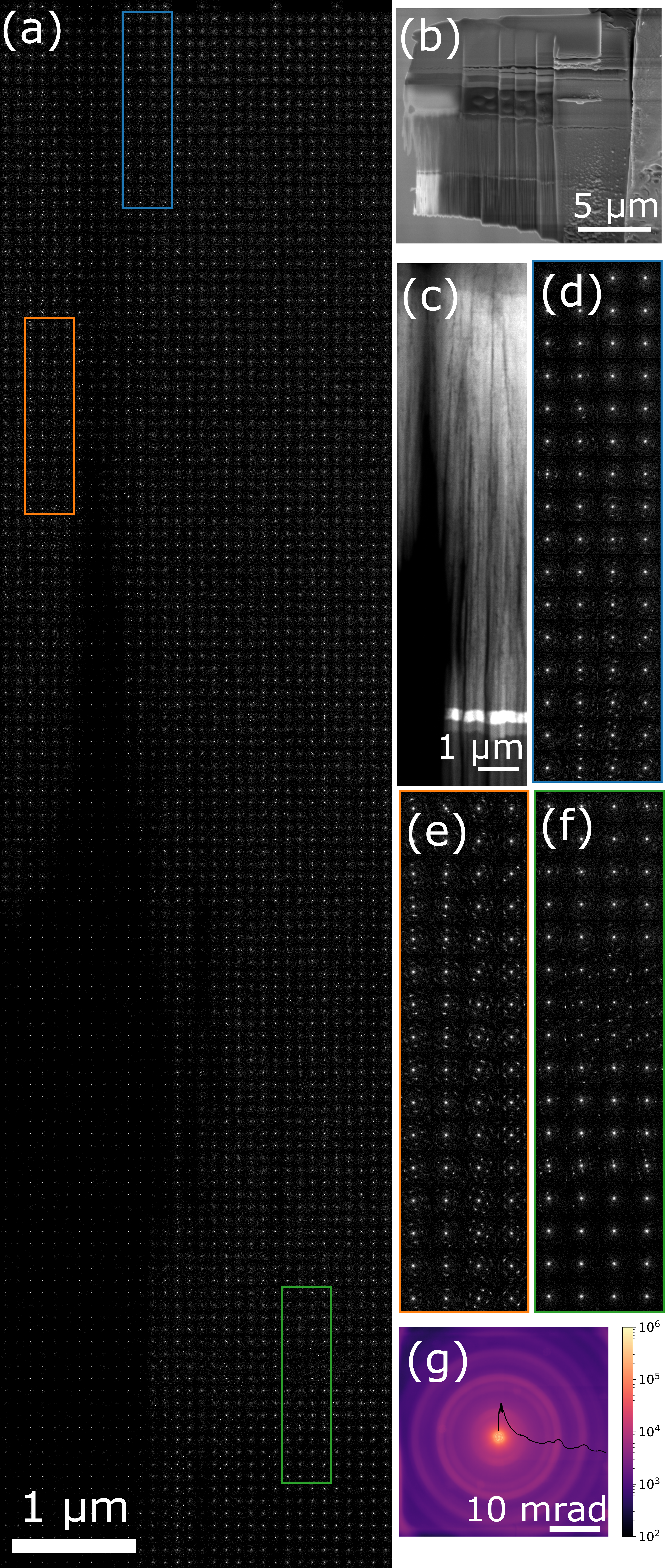}
    \caption[battery1]{Large-scale parallel beam 4D-STEM scan of a battery stack. (a) A subset of diffraction patterns tiled next to each other representing 1/16$^{th}$ of the probe positions in each direction. The diffraction patterns are binned by eight to fit the page. (b) An overview image of the full FIB liftout sample. The left hand side is sufficiently thin for 4D-STEM measurements. (c) The simultaneously acquired ADF-STEM image for one 4D-STEM scan showing the full battery stack with Z-contrast. The regions in the (d) blue, (e) orange, and (f) green boxes shown in (a) are enlarged to show the details of the diffraction patterns for different regions of interest. (g) The sum of all diffraction patterns overlaid with the radial sum of all diffraction patterns showing several diffraction rings. The colorbar is in units of electrons and on a log scale.}
    \label{battery1}
\end{figure*}

The ability of the detector to investigate an entire device was tested on a focused ion beam (FIB) lift-out of a 15 by 15 $\mu$m sized battery stack composed of (from top to bottom) platinum and aluminum protective layers, titanium, LiPON, LiCoO$_2$, nickel, aluminum and silicon as shown in the overview SEM image of the entire FIB liftout seen in Figure \ref{battery1}(b). The thickness of each region of the battery stack can be differentiated by Z-contrast in the ADF-STEM image in Figure \ref{battery1}(c) simultaneously acquired with the 4D-STEM data.

The 4D Camera was used to acquire several separate scans of the thin area of the specimen with 256 by 1024 real space probe positions to improve the signal to noise. In total ten frames were acquired per probe position with a total acquisition time of 38 seconds. The rapidity of the scans and the nanometer scale scanning step size meant that sample drift could be ignored. The microscope was operated at 300 kV and configured in a parallel beam configuration with a convergence angle of about 1 mrad. A patterned aperture was used to improve the ability to localize individual diffraction spots.\cite{zeltmann_patterned_2020} The beam was rastered with a step size of 12.8 nm over a field-of-view of 3.27 by 13.07 $\mu$m. Each scan was reduced to electron counts, and the shifts of the electron beam on the detector due to scanning a large field of view were removed using shifts from a vacuum scan acquired at the same STEM magnification and camera length. The reduced, combined data set (originally 1,740 GB of raw data) only required 7 GB of disk space in the sparse format.

\begin{figure*}
    \centering
   \includegraphics[width=2.8in]{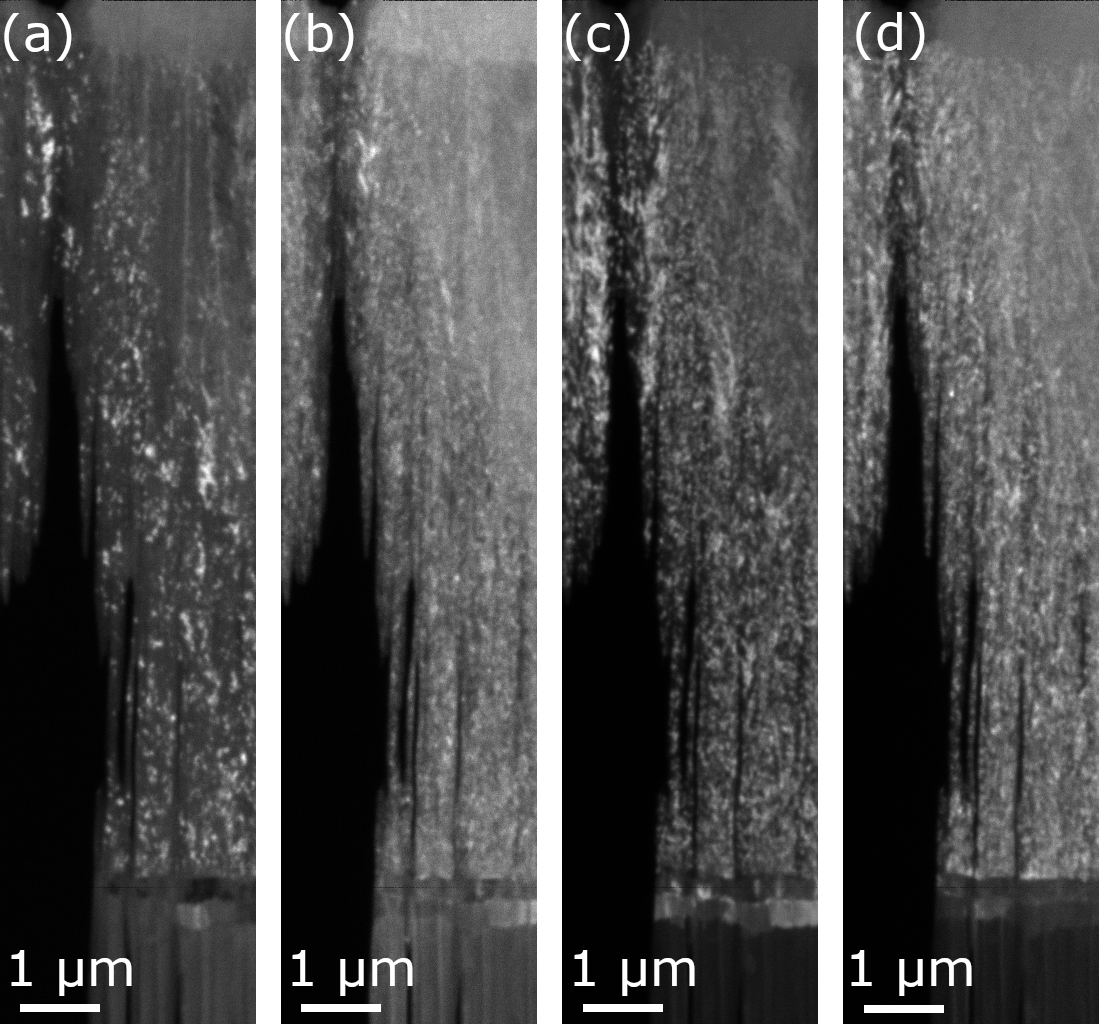}
    \caption[battery2]{vADF-STEM images generated using a rotational sum for each diffraction pattern. Refer to Figure \ref{battery1}(g) to see each ring in the full data set. The images are generated using scattering from the (a) $[111]$, (b) $[311]$ and $[222]$, (c) $[400]$, and (d) $[440]$ family of reflections in LiCO$_2$.}
    \label{battery2}
\end{figure*}

A tiled visualization of the 4D-STEM data is shown in Figure \ref{battery1}(a) along with the simultaneously acquired HAADF-STEM from only the first scan (Figure \ref{battery1}(c). To fit the page, only one in sixteen patterns is displayed, and each diffraction pattern is binned by eight to 72 by 72 pixels. The regions of interest (ROIs) enclosed in blue, orange, and green boxes are enlarged to show more detail. Figure \ref{battery1}(d) is from the blue rectangle where the LiPON directly contacts the LiCoO$_2$ for the electrolyte-electrode interface. It is clear that the LiPON is amorphous and that the interface is rough. The orange ROI enlarged in \ref{battery1}(e) is from a thin LiCoO$_2$ region and shows that diffraction patterns along the vertical direction are more similar as compared to their horizontal neighbors indicating grains are extended vertically along the growth direction (columnar growth). The patterns in the green rectangle (Figure \ref{battery1}(f)) are from the LiCoO$_2$ / nickel / aluminum interfaces at the bottom electrode. Differences in crystal orientation and structure are clearly visible, but require advanced processing for quantitative understanding of such a large data set beyond the scope of this demonstration. \cite{rauch_rapid_2005, brunetti_confirmation_2011, savitzky_py4dstem_2021, Ophus2022-zp}

Figure \ref{battery1}(g) is the sum of every pattern in the full data set showing several distinct rings as well as the shape of the patterned aperture. Interestingly, the inner-most ring is brighter on the top and bottom compared to the horizontal direction which matches the real space horizontal-vertical scanning directions. This supports a preferential direction for the LiCoO$_2$ crystals visually identified in \ref{battery1}(e). Summation of intensity for a range of scattering angles for rings of interest produces the vADF-STEM images shown in Figure \ref{battery2}. The images correspond to scattering from the LiCo$_2$ (a) $[111]$, (b) $[311]$ and $[222]$, (c) $[400]$, and (d) $[440]$ family of reflections. \cite{Kushida2002-wb, Wang1999-pp}

Solid state battery degradation and failure are directly linked to the electrochemical and mechanical limitations of bulk electrode, electrolyte material layers and their interfaces. \cite{Tan2020-mz} Data acquisition at this scale and rapidity provides bulk and localized, granular structural information of the solid state battery system that can be used to highlight crystallographic preferences within bulk LiCoO$_2$, modifications in structure of LiCoO$_2$ from electrode bulk to electrode-electrolyte interface, as well as other structural patterns. This data, if combined with complementary EELS information, can develop a statistically robust analysis of initial microstructure, charge-discharge microstructure evolution as affected by lithium ion movement, and finally, breakdown of lithium ion transport, i.e., battery failure.

\section{Summary}

The 4D Camera's high frame rate and data rate enables rapid scanning with the possibility to integrate 4D-STEM into more complex experiments and modalities. The integration of the camera with large computation resources improves feedback time to the operator from days to minutes. The 4D Camera allows a STEM to be treated as a multi-functional electron scattering beamline which goes beyond the traditional ``single image acquisition" approach in favor of capturing all scattering information where one can effectively perform many experiments on a single, large multi-dimensional data set with massive data analytics.

\section{Acknowledgements}

The experiments were performed at the Molecular Foundry, Lawrence Berkeley National Laboratory, which is supported by the U.S. Department of Energy under contract no. DE-AC02-05CH11231. Development of the 4D Camera was supported by the US Department of Energy, Office of Basic Energy Sciences, Accelerator and Detector Research Program. This work was partially funded by the US Department of Energy in the program "4D Camera Distillery: From Massive Electron Microscopy Scattering Data to Useful Information with AI/ML." We greatly appreciate the contributions of Gatan, Inc. in the development, installation, and ongoing support of the 4D Camera. This research used resources of the National Energy Research Scientific Computing Center (NERSC), a U.S. Department of Energy Office of Science User Facility located at Lawrence Berkeley National Laboratory, operated under Contract No. DE-AC02-05CH11231 using NERSC award ERCAP0020898 and ERCAP0020897. 

\newpage

\section{References}
\bibliography{4Dcamera}

\begin{thebibliography}{56}%
\makeatletter
\providecommand \@ifxundefined [1]{%
 \@ifx{#1\undefined}
}%
\providecommand \@ifnum [1]{%
 \ifnum #1\expandafter \@firstoftwo
 \else \expandafter \@secondoftwo
 \fi
}%
\providecommand \@ifx [1]{%
 \ifx #1\expandafter \@firstoftwo
 \else \expandafter \@secondoftwo
 \fi
}%
\providecommand \natexlab [1]{#1}%
\providecommand \enquote  [1]{``#1''}%
\providecommand \bibnamefont  [1]{#1}%
\providecommand \bibfnamefont [1]{#1}%
\providecommand \citenamefont [1]{#1}%
\providecommand \href@noop [0]{\@secondoftwo}%
\providecommand \href [0]{\begingroup \@sanitize@url \@href}%
\providecommand \@href[1]{\@@startlink{#1}\@@href}%
\providecommand \@@href[1]{\endgroup#1\@@endlink}%
\providecommand \@sanitize@url [0]{\catcode `\\12\catcode `\$12\catcode
  `\&12\catcode `\#12\catcode `\^12\catcode `\_12\catcode `\%12\relax}%
\providecommand \@@startlink[1]{}%
\providecommand \@@endlink[0]{}%
\providecommand \url  [0]{\begingroup\@sanitize@url \@url }%
\providecommand \@url [1]{\endgroup\@href {#1}{\urlprefix }}%
\providecommand \urlprefix  [0]{URL }%
\providecommand \Eprint [0]{\href }%
\providecommand \doibase [0]{http://dx.doi.org/}%
\providecommand \selectlanguage [0]{\@gobble}%
\providecommand \bibinfo  [0]{\@secondoftwo}%
\providecommand \bibfield  [0]{\@secondoftwo}%
\providecommand \translation [1]{[#1]}%
\providecommand \BibitemOpen [0]{}%
\providecommand \bibitemStop [0]{}%
\providecommand \bibitemNoStop [0]{.\EOS\space}%
\providecommand \EOS [0]{\spacefactor3000\relax}%
\providecommand \BibitemShut  [1]{\csname bibitem#1\endcsname}%
\let\auto@bib@innerbib\@empty
\bibitem [{\citenamefont {McMullan}\ \emph {et~al.}(2009)\citenamefont
  {McMullan}, \citenamefont {Chen}, \citenamefont {Henderson},\ and\
  \citenamefont {Faruqi}}]{mcmullan_detective_2009}%
  \BibitemOpen
  \bibfield  {author} {\bibinfo {author} {\bibfnamefont {G.}~\bibnamefont
  {McMullan}}, \bibinfo {author} {\bibfnamefont {S.}~\bibnamefont {Chen}},
  \bibinfo {author} {\bibfnamefont {R.}~\bibnamefont {Henderson}}, \ and\
  \bibinfo {author} {\bibfnamefont {A.~R.}\ \bibnamefont {Faruqi}},\ }\href
  {\doibase 10.1016/j.ultramic.2009.04.002} {\bibfield  {journal} {\bibinfo
  {journal} {Ultramicroscopy}\ }\textbf {\bibinfo {volume} {109}},\ \bibinfo
  {pages} {1126} (\bibinfo {year} {2009})}\BibitemShut {NoStop}%
\bibitem [{\citenamefont {Grob}\ \emph {et~al.}(2013)\citenamefont {Grob},
  \citenamefont {Bean}, \citenamefont {Typke}, \citenamefont {Li},
  \citenamefont {Nogales},\ and\ \citenamefont {Glaeser}}]{grob_ranking_2013}%
  \BibitemOpen
  \bibfield  {author} {\bibinfo {author} {\bibfnamefont {P.}~\bibnamefont
  {Grob}}, \bibinfo {author} {\bibfnamefont {D.}~\bibnamefont {Bean}}, \bibinfo
  {author} {\bibfnamefont {D.}~\bibnamefont {Typke}}, \bibinfo {author}
  {\bibfnamefont {X.}~\bibnamefont {Li}}, \bibinfo {author} {\bibfnamefont
  {E.}~\bibnamefont {Nogales}}, \ and\ \bibinfo {author} {\bibfnamefont
  {R.~M.}\ \bibnamefont {Glaeser}},\ }\href {\doibase
  10.1016/j.ultramic.2013.01.003} {\bibfield  {journal} {\bibinfo  {journal}
  {Ultramicroscopy}\ }\textbf {\bibinfo {volume} {133}},\ \bibinfo {pages} {1}
  (\bibinfo {year} {2013})}\BibitemShut {NoStop}%
\bibitem [{\citenamefont {Brilot}\ \emph {et~al.}(2012)\citenamefont {Brilot},
  \citenamefont {Chen}, \citenamefont {Cheng}, \citenamefont {Pan},
  \citenamefont {Harrison}, \citenamefont {Potter}, \citenamefont {Carragher},
  \citenamefont {Henderson},\ and\ \citenamefont
  {Grigorieff}}]{brilot_beam-induced_2012}%
  \BibitemOpen
  \bibfield  {author} {\bibinfo {author} {\bibfnamefont {A.~F.}\ \bibnamefont
  {Brilot}}, \bibinfo {author} {\bibfnamefont {J.~Z.}\ \bibnamefont {Chen}},
  \bibinfo {author} {\bibfnamefont {A.}~\bibnamefont {Cheng}}, \bibinfo
  {author} {\bibfnamefont {J.}~\bibnamefont {Pan}}, \bibinfo {author}
  {\bibfnamefont {S.~C.}\ \bibnamefont {Harrison}}, \bibinfo {author}
  {\bibfnamefont {C.~S.}\ \bibnamefont {Potter}}, \bibinfo {author}
  {\bibfnamefont {B.}~\bibnamefont {Carragher}}, \bibinfo {author}
  {\bibfnamefont {R.}~\bibnamefont {Henderson}}, \ and\ \bibinfo {author}
  {\bibfnamefont {N.}~\bibnamefont {Grigorieff}},\ }\href {\doibase
  10.1016/j.jsb.2012.02.003} {\bibfield  {journal} {\bibinfo  {journal}
  {Journal of Structural Biology}\ }\textbf {\bibinfo {volume} {177}},\
  \bibinfo {pages} {630} (\bibinfo {year} {2012})}\BibitemShut {NoStop}%
\bibitem [{\citenamefont {Park}\ \emph {et~al.}(2015)\citenamefont {Park},
  \citenamefont {Elmlund}, \citenamefont {Ercius}, \citenamefont {Yuk},
  \citenamefont {Limmer}, \citenamefont {Chen}, \citenamefont {Kim},
  \citenamefont {Han}, \citenamefont {Weitz}, \citenamefont {Zettl},\ and\
  \citenamefont {Alivisatos}}]{park_3d_2015}%
  \BibitemOpen
  \bibfield  {author} {\bibinfo {author} {\bibfnamefont {J.}~\bibnamefont
  {Park}}, \bibinfo {author} {\bibfnamefont {H.}~\bibnamefont {Elmlund}},
  \bibinfo {author} {\bibfnamefont {P.}~\bibnamefont {Ercius}}, \bibinfo
  {author} {\bibfnamefont {J.~M.}\ \bibnamefont {Yuk}}, \bibinfo {author}
  {\bibfnamefont {D.~T.}\ \bibnamefont {Limmer}}, \bibinfo {author}
  {\bibfnamefont {Q.}~\bibnamefont {Chen}}, \bibinfo {author} {\bibfnamefont
  {K.}~\bibnamefont {Kim}}, \bibinfo {author} {\bibfnamefont {S.~H.}\
  \bibnamefont {Han}}, \bibinfo {author} {\bibfnamefont {D.~A.}\ \bibnamefont
  {Weitz}}, \bibinfo {author} {\bibfnamefont {A.}~\bibnamefont {Zettl}}, \ and\
  \bibinfo {author} {\bibfnamefont {A.~P.}\ \bibnamefont {Alivisatos}},\ }\href
  {\doibase 10.1126/science.aab1343} {\bibfield  {journal} {\bibinfo  {journal}
  {Science}\ }\textbf {\bibinfo {volume} {349}},\ \bibinfo {pages} {290}
  (\bibinfo {year} {2015})}\BibitemShut {NoStop}%
\bibitem [{\citenamefont {Kim}\ \emph {et~al.}(2020)\citenamefont {Kim},
  \citenamefont {Heo}, \citenamefont {Kim}, \citenamefont {Reboul},
  \citenamefont {Chun}, \citenamefont {Kang}, \citenamefont {Bae},
  \citenamefont {Hyun}, \citenamefont {Lim}, \citenamefont {Lee}, \citenamefont
  {Han}, \citenamefont {Hyeon}, \citenamefont {Alivisatos}, \citenamefont
  {Ercius}, \citenamefont {Elmlund},\ and\ \citenamefont
  {Park}}]{kim_critical_2020}%
  \BibitemOpen
  \bibfield  {author} {\bibinfo {author} {\bibfnamefont {B.~H.}\ \bibnamefont
  {Kim}}, \bibinfo {author} {\bibfnamefont {J.}~\bibnamefont {Heo}}, \bibinfo
  {author} {\bibfnamefont {S.}~\bibnamefont {Kim}}, \bibinfo {author}
  {\bibfnamefont {C.~F.}\ \bibnamefont {Reboul}}, \bibinfo {author}
  {\bibfnamefont {H.}~\bibnamefont {Chun}}, \bibinfo {author} {\bibfnamefont
  {D.}~\bibnamefont {Kang}}, \bibinfo {author} {\bibfnamefont {H.}~\bibnamefont
  {Bae}}, \bibinfo {author} {\bibfnamefont {H.}~\bibnamefont {Hyun}}, \bibinfo
  {author} {\bibfnamefont {J.}~\bibnamefont {Lim}}, \bibinfo {author}
  {\bibfnamefont {H.}~\bibnamefont {Lee}}, \bibinfo {author} {\bibfnamefont
  {B.}~\bibnamefont {Han}}, \bibinfo {author} {\bibfnamefont {T.}~\bibnamefont
  {Hyeon}}, \bibinfo {author} {\bibfnamefont {A.~P.}\ \bibnamefont
  {Alivisatos}}, \bibinfo {author} {\bibfnamefont {P.}~\bibnamefont {Ercius}},
  \bibinfo {author} {\bibfnamefont {H.}~\bibnamefont {Elmlund}}, \ and\
  \bibinfo {author} {\bibfnamefont {J.}~\bibnamefont {Park}},\ }\href {\doibase
  10.1126/science.aax3233} {\bibfield  {journal} {\bibinfo  {journal}
  {Science}\ }\textbf {\bibinfo {volume} {368}},\ \bibinfo {pages} {60}
  (\bibinfo {year} {2020})}\BibitemShut {NoStop}%
\bibitem [{\citenamefont {Hart}\ \emph {et~al.}(2017)\citenamefont {Hart},
  \citenamefont {Lang}, \citenamefont {Leff}, \citenamefont {Longo},
  \citenamefont {Trevor}, \citenamefont {Twesten},\ and\ \citenamefont
  {Taheri}}]{hart_direct_2017}%
  \BibitemOpen
  \bibfield  {author} {\bibinfo {author} {\bibfnamefont {J.~L.}\ \bibnamefont
  {Hart}}, \bibinfo {author} {\bibfnamefont {A.~C.}\ \bibnamefont {Lang}},
  \bibinfo {author} {\bibfnamefont {A.~C.}\ \bibnamefont {Leff}}, \bibinfo
  {author} {\bibfnamefont {P.}~\bibnamefont {Longo}}, \bibinfo {author}
  {\bibfnamefont {C.}~\bibnamefont {Trevor}}, \bibinfo {author} {\bibfnamefont
  {R.~D.}\ \bibnamefont {Twesten}}, \ and\ \bibinfo {author} {\bibfnamefont
  {M.~L.}\ \bibnamefont {Taheri}},\ }\href {\doibase
  10.1038/s41598-017-07709-4} {\bibfield  {journal} {\bibinfo  {journal}
  {Scientific Reports}\ }\textbf {\bibinfo {volume} {7}},\ \bibinfo {pages}
  {8243} (\bibinfo {year} {2017})}\BibitemShut {NoStop}%
\bibitem [{\citenamefont {Susarla}\ \emph {et~al.}(2022)\citenamefont
  {Susarla}, \citenamefont {Naik}, \citenamefont {Blach}, \citenamefont
  {Zipfel}, \citenamefont {Taniguchi}, \citenamefont {Watanabe}, \citenamefont
  {Huang}, \citenamefont {Ramesh}, \citenamefont {da~Jornada}, \citenamefont
  {Louie}, \citenamefont {Ercius},\ and\ \citenamefont
  {Raja}}]{Susarla2022-ed}%
  \BibitemOpen
  \bibfield  {author} {\bibinfo {author} {\bibfnamefont {S.}~\bibnamefont
  {Susarla}}, \bibinfo {author} {\bibfnamefont {M.~H.}\ \bibnamefont {Naik}},
  \bibinfo {author} {\bibfnamefont {D.~D.}\ \bibnamefont {Blach}}, \bibinfo
  {author} {\bibfnamefont {J.}~\bibnamefont {Zipfel}}, \bibinfo {author}
  {\bibfnamefont {T.}~\bibnamefont {Taniguchi}}, \bibinfo {author}
  {\bibfnamefont {K.}~\bibnamefont {Watanabe}}, \bibinfo {author}
  {\bibfnamefont {L.}~\bibnamefont {Huang}}, \bibinfo {author} {\bibfnamefont
  {R.}~\bibnamefont {Ramesh}}, \bibinfo {author} {\bibfnamefont {F.~H.}\
  \bibnamefont {da~Jornada}}, \bibinfo {author} {\bibfnamefont {S.~G.}\
  \bibnamefont {Louie}}, \bibinfo {author} {\bibfnamefont {P.}~\bibnamefont
  {Ercius}}, \ and\ \bibinfo {author} {\bibfnamefont {A.}~\bibnamefont
  {Raja}},\ }\href {\doibase 10.1126/science.add9294} {\bibfield  {journal}
  {\bibinfo  {journal} {Science}\ }\textbf {\bibinfo {volume} {378}},\ \bibinfo
  {pages} {1235} (\bibinfo {year} {2022})}\BibitemShut {NoStop}%
\bibitem [{\citenamefont {Ophus}(2019)}]{ophus_four-dimensional_2019}%
  \BibitemOpen
  \bibfield  {author} {\bibinfo {author} {\bibfnamefont {C.}~\bibnamefont
  {Ophus}},\ }\href {\doibase 10.1017/S1431927619000497} {\bibfield  {journal}
  {\bibinfo  {journal} {Microscopy and Microanalysis}\ }\textbf {\bibinfo
  {volume} {25}},\ \bibinfo {pages} {563} (\bibinfo {year} {2019})}\BibitemShut
  {NoStop}%
\bibitem [{\citenamefont {Nord}\ \emph {et~al.}(2020)\citenamefont {Nord},
  \citenamefont {Webster}, \citenamefont {Paton}, \citenamefont {McVitie},
  \citenamefont {McGrouther}, \citenamefont {MacLaren},\ and\ \citenamefont
  {Paterson}}]{nord2020fast}%
  \BibitemOpen
  \bibfield  {author} {\bibinfo {author} {\bibfnamefont {M.}~\bibnamefont
  {Nord}}, \bibinfo {author} {\bibfnamefont {R.~W.}\ \bibnamefont {Webster}},
  \bibinfo {author} {\bibfnamefont {K.~A.}\ \bibnamefont {Paton}}, \bibinfo
  {author} {\bibfnamefont {S.}~\bibnamefont {McVitie}}, \bibinfo {author}
  {\bibfnamefont {D.}~\bibnamefont {McGrouther}}, \bibinfo {author}
  {\bibfnamefont {I.}~\bibnamefont {MacLaren}}, \ and\ \bibinfo {author}
  {\bibfnamefont {G.~W.}\ \bibnamefont {Paterson}},\ }\href@noop {} {\bibfield
  {journal} {\bibinfo  {journal} {Microscopy and Microanalysis}\ }\textbf
  {\bibinfo {volume} {26}},\ \bibinfo {pages} {653} (\bibinfo {year}
  {2020})}\BibitemShut {NoStop}%
\bibitem [{\citenamefont {Paterson}\ \emph {et~al.}(2020)\citenamefont
  {Paterson}, \citenamefont {Webster}, \citenamefont {Ross}, \citenamefont
  {Paton}, \citenamefont {Macgregor}, \citenamefont {McGrouther}, \citenamefont
  {MacLaren},\ and\ \citenamefont {Nord}}]{paterson2020fast}%
  \BibitemOpen
  \bibfield  {author} {\bibinfo {author} {\bibfnamefont {G.~W.}\ \bibnamefont
  {Paterson}}, \bibinfo {author} {\bibfnamefont {R.~W.}\ \bibnamefont
  {Webster}}, \bibinfo {author} {\bibfnamefont {A.}~\bibnamefont {Ross}},
  \bibinfo {author} {\bibfnamefont {K.~A.}\ \bibnamefont {Paton}}, \bibinfo
  {author} {\bibfnamefont {T.~A.}\ \bibnamefont {Macgregor}}, \bibinfo {author}
  {\bibfnamefont {D.}~\bibnamefont {McGrouther}}, \bibinfo {author}
  {\bibfnamefont {I.}~\bibnamefont {MacLaren}}, \ and\ \bibinfo {author}
  {\bibfnamefont {M.}~\bibnamefont {Nord}},\ }\href@noop {} {\bibfield
  {journal} {\bibinfo  {journal} {Microscopy and Microanalysis}\ }\textbf
  {\bibinfo {volume} {26}},\ \bibinfo {pages} {944} (\bibinfo {year}
  {2020})}\BibitemShut {NoStop}%
\bibitem [{\citenamefont {Chen}\ \emph {et~al.}(2021)\citenamefont {Chen},
  \citenamefont {Jiang}, \citenamefont {Shao}, \citenamefont {Holtz},
  \citenamefont {Odstrcil}, \citenamefont {Guizar-Sicairos}, \citenamefont
  {Hanke}, \citenamefont {Ganschow}, \citenamefont {Schlom},\ and\
  \citenamefont {Muller}}]{chen_electron_2021}%
  \BibitemOpen
  \bibfield  {author} {\bibinfo {author} {\bibfnamefont {Z.}~\bibnamefont
  {Chen}}, \bibinfo {author} {\bibfnamefont {Y.}~\bibnamefont {Jiang}},
  \bibinfo {author} {\bibfnamefont {Y.-T.}\ \bibnamefont {Shao}}, \bibinfo
  {author} {\bibfnamefont {M.~E.}\ \bibnamefont {Holtz}}, \bibinfo {author}
  {\bibfnamefont {M.}~\bibnamefont {Odstrcil}}, \bibinfo {author}
  {\bibfnamefont {M.}~\bibnamefont {Guizar-Sicairos}}, \bibinfo {author}
  {\bibfnamefont {I.}~\bibnamefont {Hanke}}, \bibinfo {author} {\bibfnamefont
  {S.}~\bibnamefont {Ganschow}}, \bibinfo {author} {\bibfnamefont {D.~G.}\
  \bibnamefont {Schlom}}, \ and\ \bibinfo {author} {\bibfnamefont {D.~A.}\
  \bibnamefont {Muller}},\ }\href {\doibase 10.1126/science.abg2533} {\bibfield
   {journal} {\bibinfo  {journal} {Science}\ }\textbf {\bibinfo {volume}
  {372}},\ \bibinfo {pages} {826} (\bibinfo {year} {2021})}\BibitemShut
  {NoStop}%
\bibitem [{\citenamefont {Yang}\ \emph {et~al.}(2017)\citenamefont {Yang},
  \citenamefont {{MacLaren}}, \citenamefont {Jones}, \citenamefont {Martinez},
  \citenamefont {Simson}, \citenamefont {Huth}, \citenamefont {Ryll},
  \citenamefont {Soltau}, \citenamefont {Sagawa}, \citenamefont {Kondo},
  \citenamefont {Ophus}, \citenamefont {Ercius}, \citenamefont {Jin},
  \citenamefont {Kovacs},\ and\ \citenamefont {Nellist}}]{yang_electron_2017}%
  \BibitemOpen
  \bibfield  {author} {\bibinfo {author} {\bibfnamefont {H.}~\bibnamefont
  {Yang}}, \bibinfo {author} {\bibfnamefont {I.}~\bibnamefont {{MacLaren}}},
  \bibinfo {author} {\bibfnamefont {L.}~\bibnamefont {Jones}}, \bibinfo
  {author} {\bibfnamefont {G.~T.}\ \bibnamefont {Martinez}}, \bibinfo {author}
  {\bibfnamefont {M.}~\bibnamefont {Simson}}, \bibinfo {author} {\bibfnamefont
  {M.}~\bibnamefont {Huth}}, \bibinfo {author} {\bibfnamefont {H.}~\bibnamefont
  {Ryll}}, \bibinfo {author} {\bibfnamefont {H.}~\bibnamefont {Soltau}},
  \bibinfo {author} {\bibfnamefont {R.}~\bibnamefont {Sagawa}}, \bibinfo
  {author} {\bibfnamefont {Y.}~\bibnamefont {Kondo}}, \bibinfo {author}
  {\bibfnamefont {C.}~\bibnamefont {Ophus}}, \bibinfo {author} {\bibfnamefont
  {P.}~\bibnamefont {Ercius}}, \bibinfo {author} {\bibfnamefont
  {L.}~\bibnamefont {Jin}}, \bibinfo {author} {\bibfnamefont {A.}~\bibnamefont
  {Kovacs}}, \ and\ \bibinfo {author} {\bibfnamefont {P.~D.}\ \bibnamefont
  {Nellist}},\ }\href {\doibase 10.1016/j.ultramic.2017.02.006} {\bibfield
  {journal} {\bibinfo  {journal} {Ultramicroscopy}\ }\textbf {\bibinfo {volume}
  {180}},\ \bibinfo {pages} {173} (\bibinfo {year} {2017})}\BibitemShut
  {NoStop}%
\bibitem [{\citenamefont {Brown}\ \emph {et~al.}(2018)\citenamefont {Brown},
  \citenamefont {Chen}, \citenamefont {Weyland}, \citenamefont {Ophus},
  \citenamefont {Ciston}, \citenamefont {Allen},\ and\ \citenamefont
  {Findlay}}]{brown_structure_2018}%
  \BibitemOpen
  \bibfield  {author} {\bibinfo {author} {\bibfnamefont {H.}~\bibnamefont
  {Brown}}, \bibinfo {author} {\bibfnamefont {Z.}~\bibnamefont {Chen}},
  \bibinfo {author} {\bibfnamefont {M.}~\bibnamefont {Weyland}}, \bibinfo
  {author} {\bibfnamefont {C.}~\bibnamefont {Ophus}}, \bibinfo {author}
  {\bibfnamefont {J.}~\bibnamefont {Ciston}}, \bibinfo {author} {\bibfnamefont
  {L.}~\bibnamefont {Allen}}, \ and\ \bibinfo {author} {\bibfnamefont
  {S.}~\bibnamefont {Findlay}},\ }\href {\doibase
  10.1103/PhysRevLett.121.266102} {\bibfield  {journal} {\bibinfo  {journal}
  {Physical Review Letters}\ }\textbf {\bibinfo {volume} {121}},\ \bibinfo
  {pages} {266102} (\bibinfo {year} {2018})}\BibitemShut {NoStop}%
\bibitem [{\citenamefont {Brown}\ \emph {et~al.}(2022)\citenamefont {Brown},
  \citenamefont {Pelz}, \citenamefont {Hsu}, \citenamefont {Zhang},
  \citenamefont {Ramesh}, \citenamefont {Inzani}, \citenamefont {Sheridan},
  \citenamefont {Griffin}, \citenamefont {Schloz}, \citenamefont {Pekin},
  \citenamefont {Koch}, \citenamefont {Findlay}, \citenamefont {Allen},
  \citenamefont {Scott}, \citenamefont {Ophus},\ and\ \citenamefont
  {Ciston}}]{Brown2022-au}%
  \BibitemOpen
  \bibfield  {author} {\bibinfo {author} {\bibfnamefont {H.~G.}\ \bibnamefont
  {Brown}}, \bibinfo {author} {\bibfnamefont {P.~M.}\ \bibnamefont {Pelz}},
  \bibinfo {author} {\bibfnamefont {S.-L.}\ \bibnamefont {Hsu}}, \bibinfo
  {author} {\bibfnamefont {Z.}~\bibnamefont {Zhang}}, \bibinfo {author}
  {\bibfnamefont {R.}~\bibnamefont {Ramesh}}, \bibinfo {author} {\bibfnamefont
  {K.}~\bibnamefont {Inzani}}, \bibinfo {author} {\bibfnamefont
  {E.}~\bibnamefont {Sheridan}}, \bibinfo {author} {\bibfnamefont {S.~M.}\
  \bibnamefont {Griffin}}, \bibinfo {author} {\bibfnamefont {M.}~\bibnamefont
  {Schloz}}, \bibinfo {author} {\bibfnamefont {T.~C.}\ \bibnamefont {Pekin}},
  \bibinfo {author} {\bibfnamefont {C.~T.}\ \bibnamefont {Koch}}, \bibinfo
  {author} {\bibfnamefont {S.~D.}\ \bibnamefont {Findlay}}, \bibinfo {author}
  {\bibfnamefont {L.~J.}\ \bibnamefont {Allen}}, \bibinfo {author}
  {\bibfnamefont {M.~C.}\ \bibnamefont {Scott}}, \bibinfo {author}
  {\bibfnamefont {C.}~\bibnamefont {Ophus}}, \ and\ \bibinfo {author}
  {\bibfnamefont {J.}~\bibnamefont {Ciston}},\ }\href {\doibase
  10.1017/S1431927622012090} {\bibfield  {journal} {\bibinfo  {journal}
  {Microscopy and microanalysis: the official journal of Microscopy Society of
  America, Microbeam Analysis Society, Microscopical Society of Canada}\
  }\textbf {\bibinfo {volume} {28}},\ \bibinfo {pages} {1632} (\bibinfo {year}
  {2022})}\BibitemShut {NoStop}%
\bibitem [{\citenamefont {Hachtel}\ \emph {et~al.}(2018)\citenamefont
  {Hachtel}, \citenamefont {Idrobo},\ and\ \citenamefont
  {Chi}}]{hachtel_sub-angstrom_2018}%
  \BibitemOpen
  \bibfield  {author} {\bibinfo {author} {\bibfnamefont {J.~A.}\ \bibnamefont
  {Hachtel}}, \bibinfo {author} {\bibfnamefont {J.~C.}\ \bibnamefont {Idrobo}},
  \ and\ \bibinfo {author} {\bibfnamefont {M.}~\bibnamefont {Chi}},\ }\href
  {\doibase 10.1186/s40679-018-0059-4} {\bibfield  {journal} {\bibinfo
  {journal} {Advanced Structural and Chemical Imaging}\ }\textbf {\bibinfo
  {volume} {4}},\ \bibinfo {pages} {10} (\bibinfo {year} {2018})}\BibitemShut
  {NoStop}%
\bibitem [{\citenamefont {Murthy}\ \emph {et~al.}(2021)\citenamefont {Murthy},
  \citenamefont {Ribet}, \citenamefont {Stanev}, \citenamefont {Liu},
  \citenamefont {Watanabe}, \citenamefont {Taniguchi}, \citenamefont {Stern},
  \citenamefont {Reis},\ and\ \citenamefont {Dravid}}]{murthy_spatial_2021}%
  \BibitemOpen
  \bibfield  {author} {\bibinfo {author} {\bibfnamefont {A.~A.}\ \bibnamefont
  {Murthy}}, \bibinfo {author} {\bibfnamefont {S.~M.}\ \bibnamefont {Ribet}},
  \bibinfo {author} {\bibfnamefont {T.~K.}\ \bibnamefont {Stanev}}, \bibinfo
  {author} {\bibfnamefont {P.}~\bibnamefont {Liu}}, \bibinfo {author}
  {\bibfnamefont {K.}~\bibnamefont {Watanabe}}, \bibinfo {author}
  {\bibfnamefont {T.}~\bibnamefont {Taniguchi}}, \bibinfo {author}
  {\bibfnamefont {N.~P.}\ \bibnamefont {Stern}}, \bibinfo {author}
  {\bibfnamefont {R.~d.}\ \bibnamefont {Reis}}, \ and\ \bibinfo {author}
  {\bibfnamefont {V.~P.}\ \bibnamefont {Dravid}},\ }\href {\doibase
  10.1021/acs.nanolett.1c01636} {\bibfield  {journal} {\bibinfo  {journal}
  {Nano Letters}\ } (\bibinfo {year} {2021}),\
  10.1021/acs.nanolett.1c01636}\BibitemShut {NoStop}%
\bibitem [{\citenamefont {Hirata}\ \emph {et~al.}(2011)\citenamefont {Hirata},
  \citenamefont {Guan}, \citenamefont {Fujita}, \citenamefont {Hirotsu},
  \citenamefont {Inoue}, \citenamefont {Yavari}, \citenamefont {Sakurai},\ and\
  \citenamefont {Chen}}]{hirata_direct_2011}%
  \BibitemOpen
  \bibfield  {author} {\bibinfo {author} {\bibfnamefont {A.}~\bibnamefont
  {Hirata}}, \bibinfo {author} {\bibfnamefont {P.}~\bibnamefont {Guan}},
  \bibinfo {author} {\bibfnamefont {T.}~\bibnamefont {Fujita}}, \bibinfo
  {author} {\bibfnamefont {Y.}~\bibnamefont {Hirotsu}}, \bibinfo {author}
  {\bibfnamefont {A.}~\bibnamefont {Inoue}}, \bibinfo {author} {\bibfnamefont
  {A.~R.}\ \bibnamefont {Yavari}}, \bibinfo {author} {\bibfnamefont
  {T.}~\bibnamefont {Sakurai}}, \ and\ \bibinfo {author} {\bibfnamefont
  {M.}~\bibnamefont {Chen}},\ }\href {\doibase 10.1038/nmat2897} {\bibfield
  {journal} {\bibinfo  {journal} {Nature Materials}\ }\textbf {\bibinfo
  {volume} {10}},\ \bibinfo {pages} {28} (\bibinfo {year} {2011})}\BibitemShut
  {NoStop}%
\bibitem [{\citenamefont {Ozdol}\ \emph {et~al.}(2015)\citenamefont {Ozdol},
  \citenamefont {Gammer}, \citenamefont {Jin}, \citenamefont {Ercius},
  \citenamefont {Ophus}, \citenamefont {Ciston},\ and\ \citenamefont
  {Minor}}]{ozdol_strain_2015}%
  \BibitemOpen
  \bibfield  {author} {\bibinfo {author} {\bibfnamefont {V.~B.}\ \bibnamefont
  {Ozdol}}, \bibinfo {author} {\bibfnamefont {C.}~\bibnamefont {Gammer}},
  \bibinfo {author} {\bibfnamefont {X.~G.}\ \bibnamefont {Jin}}, \bibinfo
  {author} {\bibfnamefont {P.}~\bibnamefont {Ercius}}, \bibinfo {author}
  {\bibfnamefont {C.}~\bibnamefont {Ophus}}, \bibinfo {author} {\bibfnamefont
  {J.}~\bibnamefont {Ciston}}, \ and\ \bibinfo {author} {\bibfnamefont {A.~M.}\
  \bibnamefont {Minor}},\ }\href {\doibase 10.1063/1.4922994} {\bibfield
  {journal} {\bibinfo  {journal} {Applied Physics Letters}\ }\textbf {\bibinfo
  {volume} {106}},\ \bibinfo {pages} {253107} (\bibinfo {year}
  {2015})}\BibitemShut {NoStop}%
\bibitem [{\citenamefont {Zeltmann}\ \emph {et~al.}(2020)\citenamefont
  {Zeltmann}, \citenamefont {Müller}, \citenamefont {Bustillo}, \citenamefont
  {Savitzky}, \citenamefont {Hughes}, \citenamefont {Minor},\ and\
  \citenamefont {Ophus}}]{zeltmann_patterned_2020}%
  \BibitemOpen
  \bibfield  {author} {\bibinfo {author} {\bibfnamefont {S.~E.}\ \bibnamefont
  {Zeltmann}}, \bibinfo {author} {\bibfnamefont {A.}~\bibnamefont {Müller}},
  \bibinfo {author} {\bibfnamefont {K.~C.}\ \bibnamefont {Bustillo}}, \bibinfo
  {author} {\bibfnamefont {B.}~\bibnamefont {Savitzky}}, \bibinfo {author}
  {\bibfnamefont {L.}~\bibnamefont {Hughes}}, \bibinfo {author} {\bibfnamefont
  {A.~M.}\ \bibnamefont {Minor}}, \ and\ \bibinfo {author} {\bibfnamefont
  {C.}~\bibnamefont {Ophus}},\ }\href {\doibase 10.1016/j.ultramic.2019.112890}
  {\bibfield  {journal} {\bibinfo  {journal} {Ultramicroscopy}\ }\textbf
  {\bibinfo {volume} {209}},\ \bibinfo {pages} {112890} (\bibinfo {year}
  {2020})}\BibitemShut {NoStop}%
\bibitem [{\citenamefont {Panova}\ \emph {et~al.}(2019)\citenamefont {Panova},
  \citenamefont {Ophus}, \citenamefont {Takacs}, \citenamefont {Bustillo},
  \citenamefont {Balhorn}, \citenamefont {Salleo}, \citenamefont {Balsara},\
  and\ \citenamefont {Minor}}]{panova_diffraction_2019}%
  \BibitemOpen
  \bibfield  {author} {\bibinfo {author} {\bibfnamefont {O.}~\bibnamefont
  {Panova}}, \bibinfo {author} {\bibfnamefont {C.}~\bibnamefont {Ophus}},
  \bibinfo {author} {\bibfnamefont {C.~J.}\ \bibnamefont {Takacs}}, \bibinfo
  {author} {\bibfnamefont {K.~C.}\ \bibnamefont {Bustillo}}, \bibinfo {author}
  {\bibfnamefont {L.}~\bibnamefont {Balhorn}}, \bibinfo {author} {\bibfnamefont
  {A.}~\bibnamefont {Salleo}}, \bibinfo {author} {\bibfnamefont
  {N.}~\bibnamefont {Balsara}}, \ and\ \bibinfo {author} {\bibfnamefont
  {A.~M.}\ \bibnamefont {Minor}},\ }\href {\doibase 10.1038/s41563-019-0387-3}
  {\bibfield  {journal} {\bibinfo  {journal} {Nature Materials}\ }\textbf
  {\bibinfo {volume} {18}},\ \bibinfo {pages} {860} (\bibinfo {year}
  {2019})}\BibitemShut {NoStop}%
\bibitem [{\citenamefont {Jones}\ \emph {et~al.}(2015)\citenamefont {Jones},
  \citenamefont {Yang}, \citenamefont {Pennycook}, \citenamefont {Marshall},
  \citenamefont {Van~Aert}, \citenamefont {Browning}, \citenamefont {Castell},\
  and\ \citenamefont {Nellist}}]{jones2015smart}%
  \BibitemOpen
  \bibfield  {author} {\bibinfo {author} {\bibfnamefont {L.}~\bibnamefont
  {Jones}}, \bibinfo {author} {\bibfnamefont {H.}~\bibnamefont {Yang}},
  \bibinfo {author} {\bibfnamefont {T.~J.}\ \bibnamefont {Pennycook}}, \bibinfo
  {author} {\bibfnamefont {M.~S.}\ \bibnamefont {Marshall}}, \bibinfo {author}
  {\bibfnamefont {S.}~\bibnamefont {Van~Aert}}, \bibinfo {author}
  {\bibfnamefont {N.~D.}\ \bibnamefont {Browning}}, \bibinfo {author}
  {\bibfnamefont {M.~R.}\ \bibnamefont {Castell}}, \ and\ \bibinfo {author}
  {\bibfnamefont {P.~D.}\ \bibnamefont {Nellist}},\ }\href@noop {} {\bibfield
  {journal} {\bibinfo  {journal} {Advanced Structural and Chemical Imaging}\
  }\textbf {\bibinfo {volume} {1}},\ \bibinfo {pages} {1} (\bibinfo {year}
  {2015})}\BibitemShut {NoStop}%
\bibitem [{\citenamefont {Ophus}\ \emph
  {et~al.}(2016{\natexlab{a}})\citenamefont {Ophus}, \citenamefont {Ciston},\
  and\ \citenamefont {Nelson}}]{ophus2016correcting}%
  \BibitemOpen
  \bibfield  {author} {\bibinfo {author} {\bibfnamefont {C.}~\bibnamefont
  {Ophus}}, \bibinfo {author} {\bibfnamefont {J.}~\bibnamefont {Ciston}}, \
  and\ \bibinfo {author} {\bibfnamefont {C.~T.}\ \bibnamefont {Nelson}},\
  }\href@noop {} {\bibfield  {journal} {\bibinfo  {journal} {Ultramicroscopy}\
  }\textbf {\bibinfo {volume} {162}},\ \bibinfo {pages} {1} (\bibinfo {year}
  {2016}{\natexlab{a}})}\BibitemShut {NoStop}%
\bibitem [{\citenamefont {O'Leary}\ \emph {et~al.}(2020)\citenamefont
  {O'Leary}, \citenamefont {Allen}, \citenamefont {Huang}, \citenamefont {Kim},
  \citenamefont {Liberti}, \citenamefont {Nellist},\ and\ \citenamefont
  {Kirkland}}]{oleary_phase_2020}%
  \BibitemOpen
  \bibfield  {author} {\bibinfo {author} {\bibfnamefont {C.~M.}\ \bibnamefont
  {O'Leary}}, \bibinfo {author} {\bibfnamefont {C.~S.}\ \bibnamefont {Allen}},
  \bibinfo {author} {\bibfnamefont {C.}~\bibnamefont {Huang}}, \bibinfo
  {author} {\bibfnamefont {J.~S.}\ \bibnamefont {Kim}}, \bibinfo {author}
  {\bibfnamefont {E.}~\bibnamefont {Liberti}}, \bibinfo {author} {\bibfnamefont
  {P.~D.}\ \bibnamefont {Nellist}}, \ and\ \bibinfo {author} {\bibfnamefont
  {A.~I.}\ \bibnamefont {Kirkland}},\ }\href {\doibase 10.1063/1.5143213}
  {\bibfield  {journal} {\bibinfo  {journal} {Applied Physics Letters}\
  }\textbf {\bibinfo {volume} {116}},\ \bibinfo {pages} {124101} (\bibinfo
  {year} {2020})}\BibitemShut {NoStop}%
\bibitem [{\citenamefont {Stroppa}\ \emph {et~al.}(2023)\citenamefont
  {Stroppa}, \citenamefont {Meffert}, \citenamefont {Hoermann}, \citenamefont
  {Zambon}, \citenamefont {Bachevskaya}, \citenamefont {Remigy}, \citenamefont
  {Schulze-Briese},\ and\ \citenamefont {Piazza}}]{Stroppa2023-vy}%
  \BibitemOpen
  \bibfield  {author} {\bibinfo {author} {\bibfnamefont {D.~G.}\ \bibnamefont
  {Stroppa}}, \bibinfo {author} {\bibfnamefont {M.}~\bibnamefont {Meffert}},
  \bibinfo {author} {\bibfnamefont {C.}~\bibnamefont {Hoermann}}, \bibinfo
  {author} {\bibfnamefont {P.}~\bibnamefont {Zambon}}, \bibinfo {author}
  {\bibfnamefont {D.}~\bibnamefont {Bachevskaya}}, \bibinfo {author}
  {\bibfnamefont {H.}~\bibnamefont {Remigy}}, \bibinfo {author} {\bibfnamefont
  {C.}~\bibnamefont {Schulze-Briese}}, \ and\ \bibinfo {author} {\bibfnamefont
  {L.}~\bibnamefont {Piazza}},\ }\href {\doibase 10.1093/mictod/qaad005}
  {\bibfield  {journal} {\bibinfo  {journal} {Microscopy today}\ }\textbf
  {\bibinfo {volume} {31}},\ \bibinfo {pages} {10} (\bibinfo {year}
  {2023})}\BibitemShut {NoStop}%
\bibitem [{\citenamefont {Avery}\ \emph {et~al.}(2022)\citenamefont {Avery},
  \citenamefont {Harris}, \citenamefont {Ercius}, \citenamefont {Genova},
  \citenamefont {Hanwell},\ and\ \citenamefont {Zhao}}]{Avery2022-if}%
  \BibitemOpen
  \bibfield  {author} {\bibinfo {author} {\bibfnamefont {P.}~\bibnamefont
  {Avery}}, \bibinfo {author} {\bibfnamefont {C.}~\bibnamefont {Harris}},
  \bibinfo {author} {\bibfnamefont {P.}~\bibnamefont {Ercius}}, \bibinfo
  {author} {\bibfnamefont {A.}~\bibnamefont {Genova}}, \bibinfo {author}
  {\bibfnamefont {M.~D.}\ \bibnamefont {Hanwell}}, \ and\ \bibinfo {author}
  {\bibfnamefont {Z.}~\bibnamefont {Zhao}},\ }\href {\doibase
  10.5281/zenodo.7083493} {\enquote {\bibinfo {title} {{OpenChemistry/stempy}:
  stempy 3.2.0},}\ } (\bibinfo {year} {2022})\BibitemShut {NoStop}%
\bibitem [{\citenamefont {Battaglia}\ \emph {et~al.}(2009)\citenamefont
  {Battaglia}, \citenamefont {Contarato}, \citenamefont {Denes},\ and\
  \citenamefont {Giubilato}}]{battaglia_cluster_2009}%
  \BibitemOpen
  \bibfield  {author} {\bibinfo {author} {\bibfnamefont {M.}~\bibnamefont
  {Battaglia}}, \bibinfo {author} {\bibfnamefont {D.}~\bibnamefont
  {Contarato}}, \bibinfo {author} {\bibfnamefont {P.}~\bibnamefont {Denes}}, \
  and\ \bibinfo {author} {\bibfnamefont {P.}~\bibnamefont {Giubilato}},\ }\href
  {\doibase 10.1016/j.nima.2009.07.017} {\bibfield  {journal} {\bibinfo
  {journal} {Nuclear Instruments and Methods in Physics Research Section A:
  Accelerators, Spectrometers, Detectors and Associated Equipment}\ }\textbf
  {\bibinfo {volume} {608}},\ \bibinfo {pages} {363 } (\bibinfo {year}
  {2009})}\BibitemShut {NoStop}%
\bibitem [{\citenamefont {Caswell}\ \emph {et~al.}(2009)\citenamefont
  {Caswell}, \citenamefont {Ercius}, \citenamefont {Tate}, \citenamefont
  {Ercan}, \citenamefont {Gruner},\ and\ \citenamefont
  {Muller}}]{caswell_high-speed_2009}%
  \BibitemOpen
  \bibfield  {author} {\bibinfo {author} {\bibfnamefont {T.}~\bibnamefont
  {Caswell}}, \bibinfo {author} {\bibfnamefont {P.}~\bibnamefont {Ercius}},
  \bibinfo {author} {\bibfnamefont {M.}~\bibnamefont {Tate}}, \bibinfo {author}
  {\bibfnamefont {A.}~\bibnamefont {Ercan}}, \bibinfo {author} {\bibfnamefont
  {S.}~\bibnamefont {Gruner}}, \ and\ \bibinfo {author} {\bibfnamefont
  {D.}~\bibnamefont {Muller}},\ }\href {\doibase
  10.1016/j.ultramic.2008.11.023} {\bibfield  {journal} {\bibinfo  {journal}
  {Ultramicroscopy}\ }\textbf {\bibinfo {volume} {109}},\ \bibinfo {pages}
  {304} (\bibinfo {year} {2009})}\BibitemShut {NoStop}%
\bibitem [{\citenamefont {Denes}\ \emph {et~al.}(2007)\citenamefont {Denes},
  \citenamefont {Bussat}, \citenamefont {Lee},\ and\ \citenamefont
  {Radmillovic}}]{denes_active_2007}%
  \BibitemOpen
  \bibfield  {author} {\bibinfo {author} {\bibfnamefont {P.}~\bibnamefont
  {Denes}}, \bibinfo {author} {\bibfnamefont {J.-M.}\ \bibnamefont {Bussat}},
  \bibinfo {author} {\bibfnamefont {Z.}~\bibnamefont {Lee}}, \ and\ \bibinfo
  {author} {\bibfnamefont {V.}~\bibnamefont {Radmillovic}},\ }\href {\doibase
  https://doi.org/10.1016/j.nima.2007.05.308} {\bibfield  {journal} {\bibinfo
  {journal} {Nuclear Instruments and Methods in Physics Research Section A:
  Accelerators, Spectrometers, Detectors and Associated Equipment}\ }\textbf
  {\bibinfo {volume} {579}},\ \bibinfo {pages} {891} (\bibinfo {year}
  {2007})}\BibitemShut {NoStop}%
\bibitem [{\citenamefont {Li}\ \emph {et~al.}(2013)\citenamefont {Li},
  \citenamefont {Zheng}, \citenamefont {Egami}, \citenamefont {Agard},\ and\
  \citenamefont {Cheng}}]{li_influence_2013}%
  \BibitemOpen
  \bibfield  {author} {\bibinfo {author} {\bibfnamefont {X.}~\bibnamefont
  {Li}}, \bibinfo {author} {\bibfnamefont {S.~Q.}\ \bibnamefont {Zheng}},
  \bibinfo {author} {\bibfnamefont {K.}~\bibnamefont {Egami}}, \bibinfo
  {author} {\bibfnamefont {D.~A.}\ \bibnamefont {Agard}}, \ and\ \bibinfo
  {author} {\bibfnamefont {Y.}~\bibnamefont {Cheng}},\ }\href {\doibase
  10.1016/j.jsb.2013.08.005} {\bibfield  {journal} {\bibinfo  {journal}
  {Journal of Structural Biology}\ }\textbf {\bibinfo {volume} {184}},\
  \bibinfo {pages} {251} (\bibinfo {year} {2013})}\BibitemShut {NoStop}%
\bibitem [{\citenamefont {Ruskin}\ \emph {et~al.}(2013)\citenamefont {Ruskin},
  \citenamefont {Yu},\ and\ \citenamefont
  {Grigorieff}}]{ruskin_quantitative_2013}%
  \BibitemOpen
  \bibfield  {author} {\bibinfo {author} {\bibfnamefont {R.~S.}\ \bibnamefont
  {Ruskin}}, \bibinfo {author} {\bibfnamefont {Z.}~\bibnamefont {Yu}}, \ and\
  \bibinfo {author} {\bibfnamefont {N.}~\bibnamefont {Grigorieff}},\ }\href
  {\doibase 10.1016/j.jsb.2013.10.016} {\bibfield  {journal} {\bibinfo
  {journal} {Journal of Structural Biology}\ }\textbf {\bibinfo {volume}
  {184}},\ \bibinfo {pages} {385} (\bibinfo {year} {2013})}\BibitemShut
  {NoStop}%
\bibitem [{\citenamefont {Pelz}\ \emph {et~al.}(2022)\citenamefont {Pelz},
  \citenamefont {Johnson}, \citenamefont {Ophus}, \citenamefont {Ercius},\ and\
  \citenamefont {Scott}}]{pelz_real-time_2022}%
  \BibitemOpen
  \bibfield  {author} {\bibinfo {author} {\bibfnamefont {P.~M.}\ \bibnamefont
  {Pelz}}, \bibinfo {author} {\bibfnamefont {I.}~\bibnamefont {Johnson}},
  \bibinfo {author} {\bibfnamefont {C.}~\bibnamefont {Ophus}}, \bibinfo
  {author} {\bibfnamefont {P.}~\bibnamefont {Ercius}}, \ and\ \bibinfo {author}
  {\bibfnamefont {M.~C.}\ \bibnamefont {Scott}},\ }\href {\doibase
  10.1109/MSP.2021.3120981} {\bibfield  {journal} {\bibinfo  {journal} {{IEEE}
  Signal Processing Magazine}\ }\textbf {\bibinfo {volume} {39}},\ \bibinfo
  {pages} {25} (\bibinfo {year} {2022})}\BibitemShut {NoStop}%
\bibitem [{\citenamefont {Bakaul}\ \emph {et~al.}(2016)\citenamefont {Bakaul},
  \citenamefont {Serrao}, \citenamefont {Lee}, \citenamefont {Yeung},
  \citenamefont {Sarker}, \citenamefont {Hsu}, \citenamefont {Yadav},
  \citenamefont {Dedon}, \citenamefont {You}, \citenamefont {Khan},
  \citenamefont {Clarkson}, \citenamefont {Hu}, \citenamefont {Ramesh},\ and\
  \citenamefont {Salahuddin}}]{bakaul_single_2016}%
  \BibitemOpen
  \bibfield  {author} {\bibinfo {author} {\bibfnamefont {S.~R.}\ \bibnamefont
  {Bakaul}}, \bibinfo {author} {\bibfnamefont {C.~R.}\ \bibnamefont {Serrao}},
  \bibinfo {author} {\bibfnamefont {M.}~\bibnamefont {Lee}}, \bibinfo {author}
  {\bibfnamefont {C.~W.}\ \bibnamefont {Yeung}}, \bibinfo {author}
  {\bibfnamefont {A.}~\bibnamefont {Sarker}}, \bibinfo {author} {\bibfnamefont
  {S.-L.}\ \bibnamefont {Hsu}}, \bibinfo {author} {\bibfnamefont {A.~K.}\
  \bibnamefont {Yadav}}, \bibinfo {author} {\bibfnamefont {L.}~\bibnamefont
  {Dedon}}, \bibinfo {author} {\bibfnamefont {L.}~\bibnamefont {You}}, \bibinfo
  {author} {\bibfnamefont {A.~I.}\ \bibnamefont {Khan}}, \bibinfo {author}
  {\bibfnamefont {J.~D.}\ \bibnamefont {Clarkson}}, \bibinfo {author}
  {\bibfnamefont {C.}~\bibnamefont {Hu}}, \bibinfo {author} {\bibfnamefont
  {R.}~\bibnamefont {Ramesh}}, \ and\ \bibinfo {author} {\bibfnamefont
  {S.}~\bibnamefont {Salahuddin}},\ }\href {\doibase 10.1038/ncomms10547}
  {\bibfield  {journal} {\bibinfo  {journal} {Nature Communications}\ }\textbf
  {\bibinfo {volume} {7}},\ \bibinfo {pages} {10547} (\bibinfo {year}
  {2016})}\BibitemShut {NoStop}%
\bibitem [{\citenamefont {LeBeau}\ \emph {et~al.}(2010)\citenamefont {LeBeau},
  \citenamefont {Findlay}, \citenamefont {Allen},\ and\ \citenamefont
  {Stemmer}}]{lebeau2010position}%
  \BibitemOpen
  \bibfield  {author} {\bibinfo {author} {\bibfnamefont {J.~M.}\ \bibnamefont
  {LeBeau}}, \bibinfo {author} {\bibfnamefont {S.~D.}\ \bibnamefont {Findlay}},
  \bibinfo {author} {\bibfnamefont {L.~J.}\ \bibnamefont {Allen}}, \ and\
  \bibinfo {author} {\bibfnamefont {S.}~\bibnamefont {Stemmer}},\ }\href@noop
  {} {\bibfield  {journal} {\bibinfo  {journal} {Ultramicroscopy}\ }\textbf
  {\bibinfo {volume} {110}},\ \bibinfo {pages} {118} (\bibinfo {year}
  {2010})}\BibitemShut {NoStop}%
\bibitem [{\citenamefont {Ophus}\ \emph {et~al.}(2017)\citenamefont {Ophus},
  \citenamefont {Ercius}, \citenamefont {Huijben},\ and\ \citenamefont
  {Ciston}}]{ophus2017nonspectroscopic}%
  \BibitemOpen
  \bibfield  {author} {\bibinfo {author} {\bibfnamefont {C.}~\bibnamefont
  {Ophus}}, \bibinfo {author} {\bibfnamefont {P.}~\bibnamefont {Ercius}},
  \bibinfo {author} {\bibfnamefont {M.}~\bibnamefont {Huijben}}, \ and\
  \bibinfo {author} {\bibfnamefont {J.}~\bibnamefont {Ciston}},\ }\href@noop {}
  {\bibfield  {journal} {\bibinfo  {journal} {Applied Physics Letters}\
  }\textbf {\bibinfo {volume} {110}},\ \bibinfo {pages} {063102} (\bibinfo
  {year} {2017})}\BibitemShut {NoStop}%
\bibitem [{\citenamefont {Ophus}\ \emph
  {et~al.}(2016{\natexlab{b}})\citenamefont {Ophus}, \citenamefont {Ciston},
  \citenamefont {Pierce}, \citenamefont {Harvey}, \citenamefont {Chess},
  \citenamefont {McMorran}, \citenamefont {Czarnik}, \citenamefont {Rose},\
  and\ \citenamefont {Ercius}}]{ophus2016efficient}%
  \BibitemOpen
  \bibfield  {author} {\bibinfo {author} {\bibfnamefont {C.}~\bibnamefont
  {Ophus}}, \bibinfo {author} {\bibfnamefont {J.}~\bibnamefont {Ciston}},
  \bibinfo {author} {\bibfnamefont {J.}~\bibnamefont {Pierce}}, \bibinfo
  {author} {\bibfnamefont {T.~R.}\ \bibnamefont {Harvey}}, \bibinfo {author}
  {\bibfnamefont {J.}~\bibnamefont {Chess}}, \bibinfo {author} {\bibfnamefont
  {B.~J.}\ \bibnamefont {McMorran}}, \bibinfo {author} {\bibfnamefont
  {C.}~\bibnamefont {Czarnik}}, \bibinfo {author} {\bibfnamefont {H.~H.}\
  \bibnamefont {Rose}}, \ and\ \bibinfo {author} {\bibfnamefont
  {P.}~\bibnamefont {Ercius}},\ }\href@noop {} {\bibfield  {journal} {\bibinfo
  {journal} {Nature communications}\ }\textbf {\bibinfo {volume} {7}},\
  \bibinfo {pages} {1} (\bibinfo {year} {2016}{\natexlab{b}})}\BibitemShut
  {NoStop}%
\bibitem [{\citenamefont {Yang}\ \emph {et~al.}(2016)\citenamefont {Yang},
  \citenamefont {Rutte}, \citenamefont {Jones}, \citenamefont {Simson},
  \citenamefont {Sagawa}, \citenamefont {Ryll}, \citenamefont {Huth},
  \citenamefont {Pennycook}, \citenamefont {Green}, \citenamefont {Soltau},
  \citenamefont {Kondo}, \citenamefont {Davis},\ and\ \citenamefont
  {Nellist}}]{yang_simultaneous_2016}%
  \BibitemOpen
  \bibfield  {author} {\bibinfo {author} {\bibfnamefont {H.}~\bibnamefont
  {Yang}}, \bibinfo {author} {\bibfnamefont {R.~N.}\ \bibnamefont {Rutte}},
  \bibinfo {author} {\bibfnamefont {L.}~\bibnamefont {Jones}}, \bibinfo
  {author} {\bibfnamefont {M.}~\bibnamefont {Simson}}, \bibinfo {author}
  {\bibfnamefont {R.}~\bibnamefont {Sagawa}}, \bibinfo {author} {\bibfnamefont
  {H.}~\bibnamefont {Ryll}}, \bibinfo {author} {\bibfnamefont {M.}~\bibnamefont
  {Huth}}, \bibinfo {author} {\bibfnamefont {T.~J.}\ \bibnamefont {Pennycook}},
  \bibinfo {author} {\bibfnamefont {M.}~\bibnamefont {Green}}, \bibinfo
  {author} {\bibfnamefont {H.}~\bibnamefont {Soltau}}, \bibinfo {author}
  {\bibfnamefont {Y.}~\bibnamefont {Kondo}}, \bibinfo {author} {\bibfnamefont
  {B.~G.}\ \bibnamefont {Davis}}, \ and\ \bibinfo {author} {\bibfnamefont
  {P.~D.}\ \bibnamefont {Nellist}},\ }\href {\doibase 10.1038/ncomms12532}
  {\bibfield  {journal} {\bibinfo  {journal} {Nature Communications}\ }\textbf
  {\bibinfo {volume} {7}},\ \bibinfo {pages} {12532} (\bibinfo {year}
  {2016})}\BibitemShut {NoStop}%
\bibitem [{\citenamefont {Shibata}\ \emph {et~al.}(2012)\citenamefont
  {Shibata}, \citenamefont {Findlay}, \citenamefont {Kohno}, \citenamefont
  {Sawada}, \citenamefont {Kondo},\ and\ \citenamefont
  {Ikuhara}}]{shibata2012differential}%
  \BibitemOpen
  \bibfield  {author} {\bibinfo {author} {\bibfnamefont {N.}~\bibnamefont
  {Shibata}}, \bibinfo {author} {\bibfnamefont {S.~D.}\ \bibnamefont
  {Findlay}}, \bibinfo {author} {\bibfnamefont {Y.}~\bibnamefont {Kohno}},
  \bibinfo {author} {\bibfnamefont {H.}~\bibnamefont {Sawada}}, \bibinfo
  {author} {\bibfnamefont {Y.}~\bibnamefont {Kondo}}, \ and\ \bibinfo {author}
  {\bibfnamefont {Y.}~\bibnamefont {Ikuhara}},\ }\href@noop {} {\bibfield
  {journal} {\bibinfo  {journal} {Nature Physics}\ }\textbf {\bibinfo {volume}
  {8}},\ \bibinfo {pages} {611} (\bibinfo {year} {2012})}\BibitemShut {NoStop}%
\bibitem [{\citenamefont {Close}\ \emph {et~al.}(2015)\citenamefont {Close},
  \citenamefont {Chen}, \citenamefont {Shibata},\ and\ \citenamefont
  {Findlay}}]{close_towards_2015}%
  \BibitemOpen
  \bibfield  {author} {\bibinfo {author} {\bibfnamefont {R.}~\bibnamefont
  {Close}}, \bibinfo {author} {\bibfnamefont {Z.}~\bibnamefont {Chen}},
  \bibinfo {author} {\bibfnamefont {N.}~\bibnamefont {Shibata}}, \ and\
  \bibinfo {author} {\bibfnamefont {S.~D.}\ \bibnamefont {Findlay}},\ }\href
  {\doibase 10.1016/j.ultramic.2015.09.002} {\bibfield  {journal} {\bibinfo
  {journal} {Ultramicroscopy}\ }\textbf {\bibinfo {volume} {159}},\ \bibinfo
  {pages} {124} (\bibinfo {year} {2015})}\BibitemShut {NoStop}%
\bibitem [{\citenamefont {Ishizuka}\ \emph {et~al.}(2017)\citenamefont
  {Ishizuka}, \citenamefont {Oka}, \citenamefont {Seki}, \citenamefont
  {Shibata},\ and\ \citenamefont
  {Ishizuka}}]{ishizuka_boundary-artifact-free_2017}%
  \BibitemOpen
  \bibfield  {author} {\bibinfo {author} {\bibfnamefont {A.}~\bibnamefont
  {Ishizuka}}, \bibinfo {author} {\bibfnamefont {M.}~\bibnamefont {Oka}},
  \bibinfo {author} {\bibfnamefont {T.}~\bibnamefont {Seki}}, \bibinfo {author}
  {\bibfnamefont {N.}~\bibnamefont {Shibata}}, \ and\ \bibinfo {author}
  {\bibfnamefont {K.}~\bibnamefont {Ishizuka}},\ }\href {\doibase
  10.1093/jmicro/dfx032} {\bibfield  {journal} {\bibinfo  {journal} {Journal of
  Electron Microscopy}\ }\textbf {\bibinfo {volume} {66}},\ \bibinfo {pages}
  {397} (\bibinfo {year} {2017})}\BibitemShut {NoStop}%
\bibitem [{\citenamefont {Cao}\ \emph {et~al.}(2018)\citenamefont {Cao},
  \citenamefont {Han}, \citenamefont {Chen}, \citenamefont {Jiang},
  \citenamefont {Nguyen}, \citenamefont {Turgut}, \citenamefont {Fuchs},\ and\
  \citenamefont {Muller}}]{cao_theory_2018}%
  \BibitemOpen
  \bibfield  {author} {\bibinfo {author} {\bibfnamefont {M.~C.}\ \bibnamefont
  {Cao}}, \bibinfo {author} {\bibfnamefont {Y.}~\bibnamefont {Han}}, \bibinfo
  {author} {\bibfnamefont {Z.}~\bibnamefont {Chen}}, \bibinfo {author}
  {\bibfnamefont {Y.}~\bibnamefont {Jiang}}, \bibinfo {author} {\bibfnamefont
  {K.~X.}\ \bibnamefont {Nguyen}}, \bibinfo {author} {\bibfnamefont
  {E.}~\bibnamefont {Turgut}}, \bibinfo {author} {\bibfnamefont {G.~D.}\
  \bibnamefont {Fuchs}}, \ and\ \bibinfo {author} {\bibfnamefont {D.~A.}\
  \bibnamefont {Muller}},\ }\href {\doibase 10.1093/jmicro/dfx123} {\bibfield
  {journal} {\bibinfo  {journal} {Microscopy}\ }\textbf {\bibinfo {volume}
  {67}},\ \bibinfo {pages} {i150} (\bibinfo {year} {2018})}\BibitemShut
  {NoStop}%
\bibitem [{\citenamefont {Yadav}\ \emph {et~al.}(2016)\citenamefont {Yadav},
  \citenamefont {Nelson}, \citenamefont {Hsu}, \citenamefont {Hong},
  \citenamefont {Clarkson}, \citenamefont {Schlepütz}, \citenamefont
  {Damodaran}, \citenamefont {Shafer}, \citenamefont {Arenholz}, \citenamefont
  {Dedon}, \citenamefont {Chen}, \citenamefont {Vishwanath}, \citenamefont
  {Minor}, \citenamefont {Chen}, \citenamefont {Scott}, \citenamefont
  {Martin},\ and\ \citenamefont {Ramesh}}]{yadav_observation_2016}%
  \BibitemOpen
  \bibfield  {author} {\bibinfo {author} {\bibfnamefont {A.~K.}\ \bibnamefont
  {Yadav}}, \bibinfo {author} {\bibfnamefont {C.~T.}\ \bibnamefont {Nelson}},
  \bibinfo {author} {\bibfnamefont {S.~L.}\ \bibnamefont {Hsu}}, \bibinfo
  {author} {\bibfnamefont {Z.}~\bibnamefont {Hong}}, \bibinfo {author}
  {\bibfnamefont {J.~D.}\ \bibnamefont {Clarkson}}, \bibinfo {author}
  {\bibfnamefont {C.~M.}\ \bibnamefont {Schlepütz}}, \bibinfo {author}
  {\bibfnamefont {A.~R.}\ \bibnamefont {Damodaran}}, \bibinfo {author}
  {\bibfnamefont {P.}~\bibnamefont {Shafer}}, \bibinfo {author} {\bibfnamefont
  {E.}~\bibnamefont {Arenholz}}, \bibinfo {author} {\bibfnamefont {L.~R.}\
  \bibnamefont {Dedon}}, \bibinfo {author} {\bibfnamefont {D.}~\bibnamefont
  {Chen}}, \bibinfo {author} {\bibfnamefont {A.}~\bibnamefont {Vishwanath}},
  \bibinfo {author} {\bibfnamefont {A.~M.}\ \bibnamefont {Minor}}, \bibinfo
  {author} {\bibfnamefont {L.~Q.}\ \bibnamefont {Chen}}, \bibinfo {author}
  {\bibfnamefont {J.~F.}\ \bibnamefont {Scott}}, \bibinfo {author}
  {\bibfnamefont {L.~W.}\ \bibnamefont {Martin}}, \ and\ \bibinfo {author}
  {\bibfnamefont {R.}~\bibnamefont {Ramesh}},\ }\href {\doibase
  10.1038/nature16463} {\bibfield  {journal} {\bibinfo  {journal} {Nature}\
  }\textbf {\bibinfo {volume} {530}},\ \bibinfo {pages} {198} (\bibinfo {year}
  {2016})}\BibitemShut {NoStop}%
\bibitem [{\citenamefont {Pedroso}\ \emph {et~al.}(2021)\citenamefont
  {Pedroso}, \citenamefont {Mann}, \citenamefont {Zuberbuhler}, \citenamefont
  {Bohn}, \citenamefont {Yu}, \citenamefont {Altoe}, \citenamefont {Craik},\
  and\ \citenamefont {Cohen}}]{Pedroso2021-wv}%
  \BibitemOpen
  \bibfield  {author} {\bibinfo {author} {\bibfnamefont {C.~C.~S.}\
  \bibnamefont {Pedroso}}, \bibinfo {author} {\bibfnamefont {V.~R.}\
  \bibnamefont {Mann}}, \bibinfo {author} {\bibfnamefont {K.}~\bibnamefont
  {Zuberbuhler}}, \bibinfo {author} {\bibfnamefont {M.-F.}\ \bibnamefont
  {Bohn}}, \bibinfo {author} {\bibfnamefont {J.}~\bibnamefont {Yu}}, \bibinfo
  {author} {\bibfnamefont {V.}~\bibnamefont {Altoe}}, \bibinfo {author}
  {\bibfnamefont {C.~S.}\ \bibnamefont {Craik}}, \ and\ \bibinfo {author}
  {\bibfnamefont {B.~E.}\ \bibnamefont {Cohen}},\ }\href {\doibase
  10.1021/acsnano.1c07856} {\bibfield  {journal} {\bibinfo  {journal} {ACS
  nano}\ }\textbf {\bibinfo {volume} {15}},\ \bibinfo {pages} {18374} (\bibinfo
  {year} {2021})}\BibitemShut {NoStop}%
\bibitem [{\citenamefont {Lee}\ \emph {et~al.}(2021)\citenamefont {Lee},
  \citenamefont {Xu}, \citenamefont {Liu}, \citenamefont {Teitelboim},
  \citenamefont {Yao}, \citenamefont {Fernandez-Bravo}, \citenamefont
  {Kotulska}, \citenamefont {Nam}, \citenamefont {Suh}, \citenamefont
  {Bednarkiewicz}, \citenamefont {Cohen}, \citenamefont {Chan},\ and\
  \citenamefont {Schuck}}]{Lee2021-rz}%
  \BibitemOpen
  \bibfield  {author} {\bibinfo {author} {\bibfnamefont {C.}~\bibnamefont
  {Lee}}, \bibinfo {author} {\bibfnamefont {E.~Z.}\ \bibnamefont {Xu}},
  \bibinfo {author} {\bibfnamefont {Y.}~\bibnamefont {Liu}}, \bibinfo {author}
  {\bibfnamefont {A.}~\bibnamefont {Teitelboim}}, \bibinfo {author}
  {\bibfnamefont {K.}~\bibnamefont {Yao}}, \bibinfo {author} {\bibfnamefont
  {A.}~\bibnamefont {Fernandez-Bravo}}, \bibinfo {author} {\bibfnamefont
  {A.~M.}\ \bibnamefont {Kotulska}}, \bibinfo {author} {\bibfnamefont {S.~H.}\
  \bibnamefont {Nam}}, \bibinfo {author} {\bibfnamefont {Y.~D.}\ \bibnamefont
  {Suh}}, \bibinfo {author} {\bibfnamefont {A.}~\bibnamefont {Bednarkiewicz}},
  \bibinfo {author} {\bibfnamefont {B.~E.}\ \bibnamefont {Cohen}}, \bibinfo
  {author} {\bibfnamefont {E.~M.}\ \bibnamefont {Chan}}, \ and\ \bibinfo
  {author} {\bibfnamefont {P.~J.}\ \bibnamefont {Schuck}},\ }\href {\doibase
  10.1038/s41586-020-03092-9} {\bibfield  {journal} {\bibinfo  {journal}
  {Nature}\ }\textbf {\bibinfo {volume} {589}},\ \bibinfo {pages} {230}
  (\bibinfo {year} {2021})}\BibitemShut {NoStop}%
\bibitem [{\citenamefont {Lee}\ \emph {et~al.}(2022)\citenamefont {Lee},
  \citenamefont {Xu}, \citenamefont {Kwock}, \citenamefont {Teitelboim},
  \citenamefont {Liu}, \citenamefont {Fardian-Melamed}, \citenamefont
  {Pedroso}, \citenamefont {Park}, \citenamefont {Kim}, \citenamefont {Pritzl},
  \citenamefont {Nam}, \citenamefont {Lohmueller}, \citenamefont {Ercius},
  \citenamefont {Suh}, \citenamefont {Cohen}, \citenamefont {Chan},\ and\
  \citenamefont {James~Schuck}}]{Lee2022-gw}%
  \BibitemOpen
  \bibfield  {author} {\bibinfo {author} {\bibfnamefont {C.}~\bibnamefont
  {Lee}}, \bibinfo {author} {\bibfnamefont {E.~Z.}\ \bibnamefont {Xu}},
  \bibinfo {author} {\bibfnamefont {K.~W.~C.}\ \bibnamefont {Kwock}}, \bibinfo
  {author} {\bibfnamefont {A.}~\bibnamefont {Teitelboim}}, \bibinfo {author}
  {\bibfnamefont {Y.}~\bibnamefont {Liu}}, \bibinfo {author} {\bibfnamefont
  {N.}~\bibnamefont {Fardian-Melamed}}, \bibinfo {author} {\bibfnamefont
  {C.~C.~S.}\ \bibnamefont {Pedroso}}, \bibinfo {author} {\bibfnamefont
  {H.~S.}\ \bibnamefont {Park}}, \bibinfo {author} {\bibfnamefont
  {J.}~\bibnamefont {Kim}}, \bibinfo {author} {\bibfnamefont {S.~D.}\
  \bibnamefont {Pritzl}}, \bibinfo {author} {\bibfnamefont {S.~H.}\
  \bibnamefont {Nam}}, \bibinfo {author} {\bibfnamefont {T.}~\bibnamefont
  {Lohmueller}}, \bibinfo {author} {\bibfnamefont {P.}~\bibnamefont {Ercius}},
  \bibinfo {author} {\bibfnamefont {Y.~D.}\ \bibnamefont {Suh}}, \bibinfo
  {author} {\bibfnamefont {B.~E.}\ \bibnamefont {Cohen}}, \bibinfo {author}
  {\bibfnamefont {E.~M.}\ \bibnamefont {Chan}}, \ and\ \bibinfo {author}
  {\bibfnamefont {P.}~\bibnamefont {James~Schuck}},\ }\href@noop {} {\
  (\bibinfo {year} {2022})},\ \Eprint {http://arxiv.org/abs/2209.06098}
  {arXiv:2209.06098 [physics.optics]} \BibitemShut {NoStop}%
\bibitem [{\citenamefont {Hobbs}(1973)}]{hobbs1973transmission}%
  \BibitemOpen
  \bibfield  {author} {\bibinfo {author} {\bibfnamefont {L.}~\bibnamefont
  {Hobbs}},\ }\href@noop {} {\bibfield  {journal} {\bibinfo  {journal} {Le
  Journal de Physique Colloques}\ }\textbf {\bibinfo {volume} {34}},\ \bibinfo
  {pages} {C9} (\bibinfo {year} {1973})}\BibitemShut {NoStop}%
\bibitem [{\citenamefont {Bosch}\ \emph {et~al.}(2016)\citenamefont {Bosch},
  \citenamefont {Lazic},\ and\ \citenamefont {Lazar}}]{bosch_integrated_2016}%
  \BibitemOpen
  \bibfield  {author} {\bibinfo {author} {\bibfnamefont {E.~G.~T.}\
  \bibnamefont {Bosch}}, \bibinfo {author} {\bibfnamefont {I.}~\bibnamefont
  {Lazic}}, \ and\ \bibinfo {author} {\bibfnamefont {S.}~\bibnamefont
  {Lazar}},\ }\href {\doibase 10.1017/S1431927616002385} {\bibfield  {journal}
  {\bibinfo  {journal} {Microscopy and Microanalysis}\ }\textbf {\bibinfo
  {volume} {22}},\ \bibinfo {pages} {306} (\bibinfo {year} {2016})}\BibitemShut
  {NoStop}%
\bibitem [{\citenamefont {Humphry}\ \emph {et~al.}(2012)\citenamefont
  {Humphry}, \citenamefont {Kraus}, \citenamefont {Hurst}, \citenamefont
  {Maiden},\ and\ \citenamefont {Rodenburg}}]{humphry_ptychographic_2012}%
  \BibitemOpen
  \bibfield  {author} {\bibinfo {author} {\bibfnamefont {M.~J.}\ \bibnamefont
  {Humphry}}, \bibinfo {author} {\bibfnamefont {B.}~\bibnamefont {Kraus}},
  \bibinfo {author} {\bibfnamefont {A.~C.}\ \bibnamefont {Hurst}}, \bibinfo
  {author} {\bibfnamefont {A.~M.}\ \bibnamefont {Maiden}}, \ and\ \bibinfo
  {author} {\bibfnamefont {J.~M.}\ \bibnamefont {Rodenburg}},\ }\href {\doibase
  10.1038/ncomms1733} {\bibfield  {journal} {\bibinfo  {journal} {Nature
  Communications}\ }\textbf {\bibinfo {volume} {3}},\ \bibinfo {pages} {730}
  (\bibinfo {year} {2012})}\BibitemShut {NoStop}%
\bibitem [{\citenamefont {Jiang}\ \emph {et~al.}(2018)\citenamefont {Jiang},
  \citenamefont {Chen}, \citenamefont {Han}, \citenamefont {Deb}, \citenamefont
  {Gao}, \citenamefont {Xie}, \citenamefont {Purohit}, \citenamefont {Tate},
  \citenamefont {Park}, \citenamefont {Gruner}, \citenamefont {Elser},\ and\
  \citenamefont {Muller}}]{jiang_electron_2018}%
  \BibitemOpen
  \bibfield  {author} {\bibinfo {author} {\bibfnamefont {Y.}~\bibnamefont
  {Jiang}}, \bibinfo {author} {\bibfnamefont {Z.}~\bibnamefont {Chen}},
  \bibinfo {author} {\bibfnamefont {Y.}~\bibnamefont {Han}}, \bibinfo {author}
  {\bibfnamefont {P.}~\bibnamefont {Deb}}, \bibinfo {author} {\bibfnamefont
  {H.}~\bibnamefont {Gao}}, \bibinfo {author} {\bibfnamefont {S.}~\bibnamefont
  {Xie}}, \bibinfo {author} {\bibfnamefont {P.}~\bibnamefont {Purohit}},
  \bibinfo {author} {\bibfnamefont {M.~W.}\ \bibnamefont {Tate}}, \bibinfo
  {author} {\bibfnamefont {J.}~\bibnamefont {Park}}, \bibinfo {author}
  {\bibfnamefont {S.~M.}\ \bibnamefont {Gruner}}, \bibinfo {author}
  {\bibfnamefont {V.}~\bibnamefont {Elser}}, \ and\ \bibinfo {author}
  {\bibfnamefont {D.~A.}\ \bibnamefont {Muller}},\ }\href {\doibase
  10.1038/s41586-018-0298-5} {\bibfield  {journal} {\bibinfo  {journal}
  {Nature}\ }\textbf {\bibinfo {volume} {559}},\ \bibinfo {pages} {343}
  (\bibinfo {year} {2018})}\BibitemShut {NoStop}%
\bibitem [{\citenamefont {Yang}\ \emph {et~al.}(2015)\citenamefont {Yang},
  \citenamefont {Pennycook},\ and\ \citenamefont
  {Nellist}}]{yang_efficient_2015}%
  \BibitemOpen
  \bibfield  {author} {\bibinfo {author} {\bibfnamefont {H.}~\bibnamefont
  {Yang}}, \bibinfo {author} {\bibfnamefont {T.~J.}\ \bibnamefont {Pennycook}},
  \ and\ \bibinfo {author} {\bibfnamefont {P.~D.}\ \bibnamefont {Nellist}},\
  }\href {\doibase 10.1016/j.ultramic.2014.10.013} {\bibfield  {journal}
  {\bibinfo  {journal} {Ultramicroscopy}\ }\bibinfo {series} {Special Issue:
  80th Birthday of Harald Rose; {PICO} 2015 – Third Conference on Frontiers
  of Aberration Corrected Electron Microscopy},\ \textbf {\bibinfo {volume}
  {151}},\ \bibinfo {pages} {232} (\bibinfo {year} {2015})}\BibitemShut
  {NoStop}%
\bibitem [{\citenamefont {Rauch}\ and\ \citenamefont
  {Dupuy}(2005)}]{rauch_rapid_2005}%
  \BibitemOpen
  \bibfield  {author} {\bibinfo {author} {\bibfnamefont {E.~F.}\ \bibnamefont
  {Rauch}}\ and\ \bibinfo {author} {\bibfnamefont {L.}~\bibnamefont {Dupuy}},\
  }\href
  {http://yadda.icm.edu.pl/baztech/element/bwmeta1.element.baztech-article-BSW3-0014-0019}
  {\bibfield  {journal} {\bibinfo  {journal} {Archives of Metallurgy and
  Materials}\ }\textbf {\bibinfo {volume} {Vol. 50, iss. 1}},\ \bibinfo {pages}
  {87} (\bibinfo {year} {2005})}\BibitemShut {NoStop}%
\bibitem [{\citenamefont {Brunetti}\ \emph {et~al.}(2011)\citenamefont
  {Brunetti}, \citenamefont {Robert}, \citenamefont {Bayle-Guillemaud},
  \citenamefont {Rouvière}, \citenamefont {Rauch}, \citenamefont {Martin},
  \citenamefont {Colin}, \citenamefont {Bertin},\ and\ \citenamefont
  {Cayron}}]{brunetti_confirmation_2011}%
  \BibitemOpen
  \bibfield  {author} {\bibinfo {author} {\bibfnamefont {G.}~\bibnamefont
  {Brunetti}}, \bibinfo {author} {\bibfnamefont {D.}~\bibnamefont {Robert}},
  \bibinfo {author} {\bibfnamefont {P.}~\bibnamefont {Bayle-Guillemaud}},
  \bibinfo {author} {\bibfnamefont {J.~L.}\ \bibnamefont {Rouvière}}, \bibinfo
  {author} {\bibfnamefont {E.~F.}\ \bibnamefont {Rauch}}, \bibinfo {author}
  {\bibfnamefont {J.~F.}\ \bibnamefont {Martin}}, \bibinfo {author}
  {\bibfnamefont {J.~F.}\ \bibnamefont {Colin}}, \bibinfo {author}
  {\bibfnamefont {F.}~\bibnamefont {Bertin}}, \ and\ \bibinfo {author}
  {\bibfnamefont {C.}~\bibnamefont {Cayron}},\ }\href {\doibase
  10.1021/cm201783z} {\bibfield  {journal} {\bibinfo  {journal} {Chemistry of
  Materials}\ }\textbf {\bibinfo {volume} {23}},\ \bibinfo {pages} {4515}
  (\bibinfo {year} {2011})}\BibitemShut {NoStop}%
\bibitem [{\citenamefont {Savitzky}\ \emph {et~al.}(2021)\citenamefont
  {Savitzky}, \citenamefont {Zeltmann}, \citenamefont {Hughes}, \citenamefont
  {Brown}, \citenamefont {Zhao}, \citenamefont {Pelz}, \citenamefont {Pekin},
  \citenamefont {Barnard}, \citenamefont {Donohue}, \citenamefont
  {Rangel~{DaCosta}}, \citenamefont {Kennedy}, \citenamefont {Xie},
  \citenamefont {Janish}, \citenamefont {Schneider}, \citenamefont {Herring},
  \citenamefont {Gopal}, \citenamefont {Anapolsky}, \citenamefont {Dhall},
  \citenamefont {Bustillo}, \citenamefont {Ercius}, \citenamefont {Scott},
  \citenamefont {Ciston}, \citenamefont {Minor},\ and\ \citenamefont
  {Ophus}}]{savitzky_py4dstem_2021}%
  \BibitemOpen
  \bibfield  {author} {\bibinfo {author} {\bibfnamefont {B.~H.}\ \bibnamefont
  {Savitzky}}, \bibinfo {author} {\bibfnamefont {S.~E.}\ \bibnamefont
  {Zeltmann}}, \bibinfo {author} {\bibfnamefont {L.~A.}\ \bibnamefont
  {Hughes}}, \bibinfo {author} {\bibfnamefont {H.~G.}\ \bibnamefont {Brown}},
  \bibinfo {author} {\bibfnamefont {S.}~\bibnamefont {Zhao}}, \bibinfo {author}
  {\bibfnamefont {P.~M.}\ \bibnamefont {Pelz}}, \bibinfo {author}
  {\bibfnamefont {T.~C.}\ \bibnamefont {Pekin}}, \bibinfo {author}
  {\bibfnamefont {E.~S.}\ \bibnamefont {Barnard}}, \bibinfo {author}
  {\bibfnamefont {J.}~\bibnamefont {Donohue}}, \bibinfo {author} {\bibfnamefont
  {L.}~\bibnamefont {Rangel~{DaCosta}}}, \bibinfo {author} {\bibfnamefont
  {E.}~\bibnamefont {Kennedy}}, \bibinfo {author} {\bibfnamefont
  {Y.}~\bibnamefont {Xie}}, \bibinfo {author} {\bibfnamefont {M.~T.}\
  \bibnamefont {Janish}}, \bibinfo {author} {\bibfnamefont {M.~M.}\
  \bibnamefont {Schneider}}, \bibinfo {author} {\bibfnamefont {P.}~\bibnamefont
  {Herring}}, \bibinfo {author} {\bibfnamefont {C.}~\bibnamefont {Gopal}},
  \bibinfo {author} {\bibfnamefont {A.}~\bibnamefont {Anapolsky}}, \bibinfo
  {author} {\bibfnamefont {R.}~\bibnamefont {Dhall}}, \bibinfo {author}
  {\bibfnamefont {K.~C.}\ \bibnamefont {Bustillo}}, \bibinfo {author}
  {\bibfnamefont {P.}~\bibnamefont {Ercius}}, \bibinfo {author} {\bibfnamefont
  {M.~C.}\ \bibnamefont {Scott}}, \bibinfo {author} {\bibfnamefont
  {J.}~\bibnamefont {Ciston}}, \bibinfo {author} {\bibfnamefont {A.~M.}\
  \bibnamefont {Minor}}, \ and\ \bibinfo {author} {\bibfnamefont
  {C.}~\bibnamefont {Ophus}},\ }\href {\doibase 10.1017/S1431927621000477}
  {\bibfield  {journal} {\bibinfo  {journal} {Microscopy and Microanalysis}\ ,\
  \bibinfo {pages} {1}} (\bibinfo {year} {2021})}\BibitemShut {NoStop}%
\bibitem [{\citenamefont {Ophus}\ \emph {et~al.}(2022)\citenamefont {Ophus},
  \citenamefont {Zeltmann}, \citenamefont {Bruefach}, \citenamefont {Rakowski},
  \citenamefont {Savitzky}, \citenamefont {Minor},\ and\ \citenamefont
  {Scott}}]{Ophus2022-zp}%
  \BibitemOpen
  \bibfield  {author} {\bibinfo {author} {\bibfnamefont {C.}~\bibnamefont
  {Ophus}}, \bibinfo {author} {\bibfnamefont {S.~E.}\ \bibnamefont {Zeltmann}},
  \bibinfo {author} {\bibfnamefont {A.}~\bibnamefont {Bruefach}}, \bibinfo
  {author} {\bibfnamefont {A.}~\bibnamefont {Rakowski}}, \bibinfo {author}
  {\bibfnamefont {B.~H.}\ \bibnamefont {Savitzky}}, \bibinfo {author}
  {\bibfnamefont {A.~M.}\ \bibnamefont {Minor}}, \ and\ \bibinfo {author}
  {\bibfnamefont {M.~C.}\ \bibnamefont {Scott}},\ }\href {\doibase
  10.1017/S1431927622000101} {\bibfield  {journal} {\bibinfo  {journal}
  {Microscopy and microanalysis: the official journal of Microscopy Society of
  America, Microbeam Analysis Society, Microscopical Society of Canada}\ ,\
  \bibinfo {pages} {1}} (\bibinfo {year} {2022})}\BibitemShut {NoStop}%
\bibitem [{\citenamefont {Kushida}\ and\ \citenamefont
  {Kuriyama}(2002)}]{Kushida2002-wb}%
  \BibitemOpen
  \bibfield  {author} {\bibinfo {author} {\bibfnamefont {K.}~\bibnamefont
  {Kushida}}\ and\ \bibinfo {author} {\bibfnamefont {K.}~\bibnamefont
  {Kuriyama}},\ }\href {\doibase 10.1016/S0038-1098(02)00325-3} {\bibfield
  {journal} {\bibinfo  {journal} {Solid state communications}\ }\textbf
  {\bibinfo {volume} {123}},\ \bibinfo {pages} {349} (\bibinfo {year}
  {2002})}\BibitemShut {NoStop}%
\bibitem [{\citenamefont {Wang}\ \emph {et~al.}(1999)\citenamefont {Wang},
  \citenamefont {Jang}, \citenamefont {Huang}, \citenamefont {Sadoway},\ and\
  \citenamefont {Chiang}}]{Wang1999-pp}%
  \BibitemOpen
  \bibfield  {author} {\bibinfo {author} {\bibfnamefont {H.}~\bibnamefont
  {Wang}}, \bibinfo {author} {\bibfnamefont {Y.}~\bibnamefont {Jang}}, \bibinfo
  {author} {\bibfnamefont {B.}~\bibnamefont {Huang}}, \bibinfo {author}
  {\bibfnamefont {D.~R.}\ \bibnamefont {Sadoway}}, \ and\ \bibinfo {author}
  {\bibfnamefont {Y.}~\bibnamefont {Chiang}},\ }\href {\doibase
  10.1149/1.1391631} {\bibfield  {journal} {\bibinfo  {journal} {Journal of the
  Electrochemical Society}\ }\textbf {\bibinfo {volume} {146}},\ \bibinfo
  {pages} {473} (\bibinfo {year} {1999})}\BibitemShut {NoStop}%
\bibitem [{\citenamefont {Tan}\ \emph {et~al.}(2020)\citenamefont {Tan},
  \citenamefont {Banerjee}, \citenamefont {Chen},\ and\ \citenamefont
  {Meng}}]{Tan2020-mz}%
  \BibitemOpen
  \bibfield  {author} {\bibinfo {author} {\bibfnamefont {D.~H.~S.}\
  \bibnamefont {Tan}}, \bibinfo {author} {\bibfnamefont {A.}~\bibnamefont
  {Banerjee}}, \bibinfo {author} {\bibfnamefont {Z.}~\bibnamefont {Chen}}, \
  and\ \bibinfo {author} {\bibfnamefont {Y.~S.}\ \bibnamefont {Meng}},\ }\href
  {\doibase 10.1038/s41565-020-0657-x} {\bibfield  {journal} {\bibinfo
  {journal} {Nature nanotechnology}\ }\textbf {\bibinfo {volume} {15}},\
  \bibinfo {pages} {170} (\bibinfo {year} {2020})}\BibitemShut {NoStop}%
\end{thebibliography}%
\end{document}